# A combined X-ray and gravitational lensing study of the massive cooling-flow cluster PKS0745-191

S.W. Allen, A.C. Fabian, J.P. Kneib
*Institute of Astronomy, Madingley Road, Cambridge CB3 OHA*

**ABSTRACT**
We present spatially-resolved X-ray spectroscopy of the X-ray luminous cluster of galaxies, PKS0745-191, using observations made with the ASCA and ROSAT satellites. The X-ray data measure the density, temperature, and metallicity profiles of the cluster gas and constrain the distribution of mass in the cluster to a radius of $\sim$ 1.5 Mpc. We report the discovery of a bright gravitationally-lensed arc at a redshift, $z_{arc} = 0.433$, in the cluster. This result identifies PKS0745 as the lowest-redshift cluster ($z_{clus} = 0.1028$) known to exhibit gravitational lensing. The properties of the lensed source are consistent with an $S_{ab}$ galaxy with ongoing star formation. The projected cluster mass determined from lensing arguments is in excellent agreement with the value measured from a multiphase analysis of the X-ray data. This contrasts with previously-reported results for the clusters Abell 1689 and Abell 2218, for which discrepancies of $\sim$ a factor 2 in the lensing and X-ray masses are observed. At both X-ray and optical wavelengths PKS0745 appears regular and dynamically relaxed. Abell 1689 and Abell 2218 exhibit more complex dynamical states, indicative of ongoing merger events. We suggest that merging activity has lead to line-of-sight mass enhancements in these systems. The X-ray spectra and images presented here confirm that PKS0745 contains one of the largest known cooling flows, with a mass deposition rate of $\sim$ 1000 $M_{\odot}$ yr$^{-1}$.

**Key words:** galaxies: clusters: individual: PKS0745-191 – cooling flows – intergalactic medium – dark matter – gravitational lensing – X-rays: galaxies

## 1   INTRODUCTION

The measurement of the distribution of mass in clusters of galaxies is of fundamental importance to modern astronomy. Clusters of galaxies are the largest gravitationally-bound systems known and knowledge of their mass provides a key probe for cosmological studies. The first measurements of the mass of clusters were based on optical observations of galaxy motions (*e.g.* Zwicky 1933. For more recent work see *e.g.* Kent & Gunn 1982; Kent & Sargent 1983; Merritt 1987). For most clusters, however, insufficient galaxy data are available to firmly constrain their mass distributions. These measurements are also complicated by substructure in the galaxy distributions, which naturally result from the growth of clusters through mergers. Recently, two further techniques for measuring the mass of clusters have been developed. These are based on X-ray observations of clusters, and observations of gravitational lensing by clusters, respectively.

Clusters of galaxies are luminous X-ray sources. The X-rays are primarily due to free-free emission from hot ($10^7 - 10^8$K) virialized gas, trapped within the deep gravitational potentials of the clusters. In the absence of forces other than thermal pressure and gravity, the hot gas (hereafter the Intracluster Medium or ICM) rapidly attains hydrostatic equilibrium with the cluster potential (Sarazin 1988). For a spherically symmetric cluster, the distribution of mass, $M(r)$, is then described by two observables; the temperature, $T$, and (electron) density, $n_e$, of the cluster gas.

$$M(r) = -\frac{rkT}{G\mu m_p} \left[ \frac{\partial \ln n_e}{\partial \ln r} + \frac{\partial \ln T}{\partial \ln r} \right] \qquad (1)$$

The first measurements of the mass of clusters from X-ray data used spectral observations made with broad-beam, collimated instruments. These instruments provided emission-weighted *average* temperatures for clusters, which, when combined with density measurements from X-ray images, allowed estimates of the total mass of the systems to be derived. However, more precise measurements of the mass distributions require the *temperature profile* of clusters to be measured and for this, spatially-resolved X-ray spectroscopy is required.

The launch of ROSAT (Trümper 1983) provided the first opportunity for detailed spatially-resolved X-ray spectroscopy of large numbers of clusters to be carried out. Although the low spectral resolution ($\sim$ 0.5 keV at 1keV) and limited observing band (0.1 − 2.4 keV) of ROSAT could only weakly constrain the temperature profiles of the hottest, most-massive clusters (*e.g.* Allen *et al.* 1993; Henry *et al.* 1993), for cooler systems (*e.g.* Centaurus: Allen & Fabian 1994) the ROSAT data provide a good measure of the emission-weighted *mean* gas temperature as a function of radius.

The spatially-resolved *single-phase* analyses possible with ROSAT (*i.e.* the determination of the properties of the intracluster gas under the assumption of a single temperature, pressure and density at each radius) provides a useful parametrization of the bulk properties of the cluster gas. However, for clusters with cooling flows - and most X-ray bright clusters contain cooling flows in the cores (Edge, Stewart & Fabian 1992) - the usefulness of this approach is limited. The presence of a cooling flow implies that the central ICM is substantially *multiphase i.e.* has a range of temperatures and densities at each particular radius. In such circumstances, the interpretation of the results from single-phase analyses is complicated.

The launch of ASCA (Tanaka, Inoue & Holt 1994) presents the first opportunity for detailed multiphase X-ray studies of clusters to be carried out. The good spectral resolution of the ASCA detectors (∼ 50eV at 0.5 keV) and the 0.4 − 12keV observing band (well-suited to the study of massive clusters) allows the spectrum of the cooling gas to be studied in greater detail than ever before. For the first time the spectrum of the cooling flow can be deconvolved from the ambient ICM, allowing the temperature profile of the ambient cluster gas (which traces the potential) to be recovered.

Since their discovery in the mid-1980s, the study of gravitational arc(let)s in clusters of galaxies has expanded to become an active area of current astronomical research. [For a recent review of this work see Fort & Mellier (1994).] These large, centrally-concentrated masses associated with clusters act like gravitational telescopes, magnifying and distorting the images of background galaxies according to the position and distance of these sources and the distribution of mass within the cluster. In clusters with multiple lensed images, the distribution of arc(lets) can be used to recover the distribution of mass in the cluster core (Mellier *et al.* 1993; Kneib *et al.* 1993). The advantage of using observations of gravitational lensing to measure the mass of clusters is that the effect is *independent of the dynamical state of the gravitating material.* No assumption of hydrostatic equilibrium is involved (unlike the X-ray and optical galaxy-dispersion methods). However, the detection of gravitationally distorted arc(let)s with ground-based telescopes is difficult and requires excellent seeing conditions. As a result, to date only a few clusters have been studied in enough detail for firm constraints on their mass distributions to be placed (Fort & Mellier 1994).

The comparison of mass measurements determined with X-ray and gravitational lensing methods allows us to examine the systematic uncertainties present in the two methods. Such a comparison provides a test of the hydrostatic assumption involved in the X-ray analyses. In addition, whereas the X-ray measurements reflect the three-dimensional potential of a cluster, the lensing measurements are (to first order) only sensitive to the *projected mass* of the system. The comparison between X-ray and lensing masses therefore probes line-of-sight structure and/or elongation in the mass distribution.

Recently Miralda-Escude & Babul (1995) and Loeb & Mao (1995) compared mass measurements from gravitational lensing and (broad-beam) X-ray techniques for two lensing clusters; Abell 1689 and Abell 2218. These authors highlighted a discrepancy of ∼ a factor two in the results obtained with the two methods, in the sense that the lensing mass exceeds the X-ray mass. They suggested that the discrepancy could indicate that the ICM in these (and possibly other) clusters is not in hydrostatic equilibrium and questioned whether non-thermal processes, such as magnetic fields or turbulent motions, could contribute substantially to the support of the cluster gas.

In this paper we describe a combined X-ray and gravitational lensing study of the X-ray luminous cluster of galaxies PKS0745-

191 (hereafter PKS0745). PKS0745 is the most-luminous cluster in the flux-limited sample of Edge *et al.* (1990) and contains one of the largest known cooling flows (Fabian *et al.* 1985; Arnaud *et al.* 1987). We report the discovery of a bright, gravitationally-lensed arc in the cluster, at $z_{arc}$ = 0.433, which identifies it as the lowest-redshift cluster ($z_{clus}$ = 0.1028) known to exhibit gravitational lensing. We present a detailed multiphase analysis of ROSAT and ASCA X-ray data and measure the mass of the cluster using both X-ray and lensing methods. The mass results show excellent agreement, in contrast to the previous results for Abell 1689 and Abell 2218. We compare the observed properties of the three clusters and show that at X-ray and optical wavelengths PKS0745 appears regular and relaxed, whereas Abell 1689 and Abell 2218 are more dynamically active. We suggest that the latter systems are currently undergoing merger events which have lead to line-of-sight enhancements in their mass, and to the discrepancies in the X-ray and lensing results.

The structure of this paper is as follows. In Section 2 we summarize the observations and the data reduction. In Section 3 we discuss the morphology of PKS0745 and comment on the unusual properties of the dominant cluster galaxy. In Section 4 we present the analysis of the X-ray spectra using both single-phase and multiphase methods. The results obtained with the two methods are compared. In Section 5 the multiphase spectral results are combined with a deprojection analysis to measure the mass of the cluster. In Section 6 we present the observations of gravitational lensing in PKS0745. We comment on the properties of the brightest lensed source and derive the cluster mass following standard lensing arguments. We compare the mass measurements derived from the X-ray and lensing data. In Section 7 we discuss the implications of our results. In Section 8 we summarize the major conclusions. Throughout this paper, we assume $H_0$=50 km s$^{-1}$ Mpc$^{-1}$, $\Omega = 1$ and $\Lambda = 0$.

## 2 OBSERVATIONS

### 2.1 The ASCA observations

A 40 ks observation of the cluster of galaxies PKS0745 was made with the ASCA X-ray satellite over the period 1993 November 6–7. The ASCA X-ray Telescope array (XRT) consists of four nested-foil telescopes, each focussed onto one of four detectors; two X-ray CCD cameras, the Solid-state Imaging Spectrometers (SIS0 and SIS1), and two Gas scintillation Imaging Spectrometers (GIS2 and GIS3). The XRT provides a spatial resolution of ∼ 3 arcmin Half Power Diameter (HPD) in the energy range 0.3 − 12 keV. The SIS detectors provide excellent spectral resolution [$\Delta E/E = 0.02(E/5.9\text{keV})^{-0.5}$] over a 22 × 22 arcmin$^2$ field of view. The GIS detectors provide poorer energy resolution [$\Delta E/E = 0.08(E/5.9\text{keV})^{-0.5}$] over a larger circular field of view of ∼ 50 arcmin diameter.

The ASCA data were reduced using the FTOOLS (version 3.0) and XSELECT (version 1.0h) software distributed by NASA Goddard Space Flight Center (GSFC). Following the standard data screening and cleaning procedures outlined in The ABC Guide to ASCA Data Reduction (version 2; ASCA Guest Observer Note, GSFC 1994) effective exposures of 33.8 ks and 32.5 ks of good data were obtained for the GIS2 and GIS3 respectively. Unfortunately, instabilities in the detector dark frame subtraction on the satellite resulted in the total loss of data for SIS1. For SIS0, however, 24.2 ks of good data were obtained.

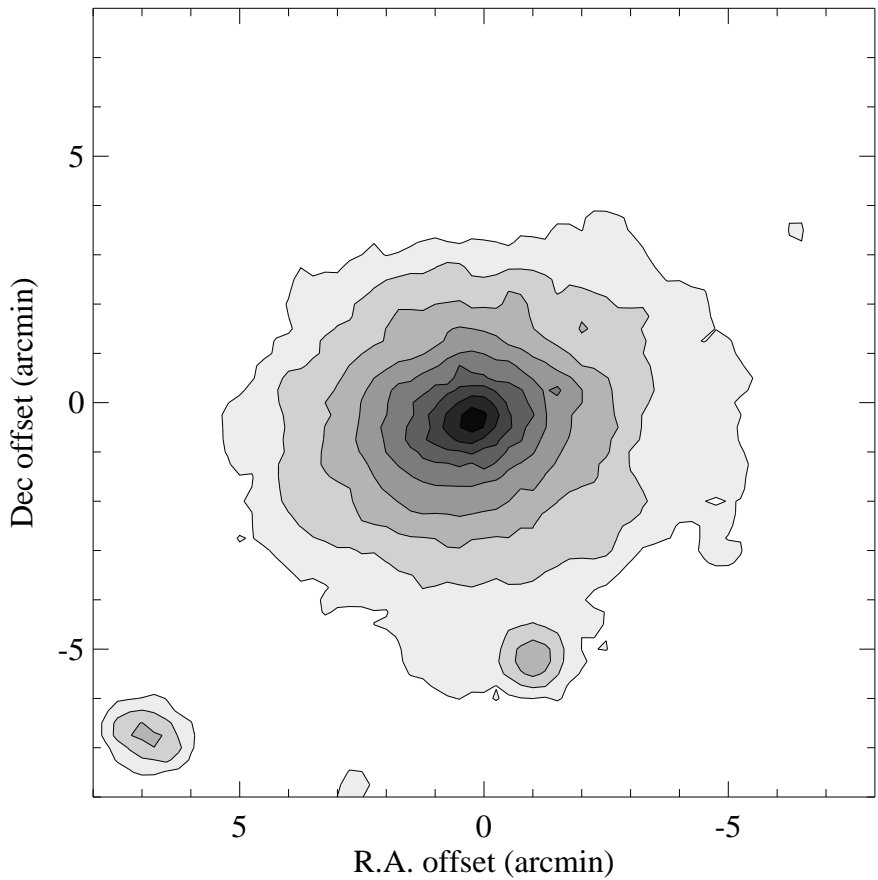

**Figure 1.** A greyscale representation of the X-ray emission from PKS0745 from the ROSAT PSPC data. The pixel size is $15 \times 15$ arcsec$^2$ and the image has been adaptively smoothed to best-illustrate the cluster morphology. The surface brightness in the raw image ranges from 0 to 338 count pixel$^{-1}$. The smoothing has been adjusted to give $\geq 16$ count pixel$^{-1}$. Contours are drawn at intervals equally separated in log space, from 0.9 to 224 count pixel$^{-1}$.

### 2.2 The ROSAT observations

Two separate observations of PKS0745 were made with the ROSAT satellite. The cluster was first observed using the High Resolution Imager (HRI) for $\sim$ 24 ks on 1992 October 20. The HRI provides a high spatial resolution ($\sim$ 4 arcsec FWHM) X-ray imaging facility in the 0.1–2.4 keV ROSAT energy band. A second 7.5 ks observation using the Position Sensitive Proportional Counter (PSPC) was made on 1993 October 15. In contrast to the HRI, the PSPC combines moderate spatial resolution ($\sim$ 25 arcsec FWHM) with a limited spectral capability [FWHM resolution $\Delta E/E = 0.43(E/0.93\text{keV})^{-0.5}$]. The PSPC also provides a higher sensitivity than the HRI for the study of extended, low-surface brightness X-ray emission.

### 2.3 The optical observations

Optical imaging of PKS0745 was carried out on 1993 December 12 with the 3.6 m Canada-France-Hawaii Telescope (CFHT), Hawaii. Exposures of 1200 s in both V and I filters were made using the Dominion Astronomical Observatory High Resolution Camera with a Loral-3 CCD detector. The Loral-3 detector has a pixel size of $0.103 \times 0.103$ arcsec$^2$ and covers a field of view of $\sim$ 2.75 $\times$ 2.75 arcmin$^2$. Flux calibration was carried out using observations of the Landolt (1992) equatorial standard star clusters SA98 and RU149. Non-photometric conditions resulted in a scatter of $\sim$ 0.2 magnitudes in the fluxes determined for the calibration sources. Seeing during the observations was $\sim$ 0.7 arcsec.

A series of optical spectra of a bright gravitational arc candidate in PKS0745 (Section 6) were obtained with the 3.5 m European Southern Observatory (ESO) New Technology Telescope (NTT) at La Silla, Chile, over the period 1994 November 27-28. Three exposures of 1200s, 1200s and 1800s (4200s total) were made using the ESO Multi-Mode Instrument (EMMI). The EMMI was used in the Red Imaging and Low Dispersion spectroscopy (RILD) mode with the 360 lines mm$^{-1}$ (No. 3) grism. A slit width of 1.5 arcsec was used and the slit was aligned with the arc, at a Position Angle (PA) of 20 degree. The adopted instrument configuration provides a spectral resolution of $\sim$ 9.4 A (at 6000A) across a wavelength range of $3900 - 8940$ A. The spectroscopic observations were carried out in seeing of $0.8 - 1.0$ arcsec.

The majority of the reduction and analysis of the optical data was carried out using the Image Reduction and Analysis Facility (IRAF). Magnitudes were determined using the SExtractor software of Bertin (1995). A full summary of all the X-ray and optical observations is provided in Table 1.

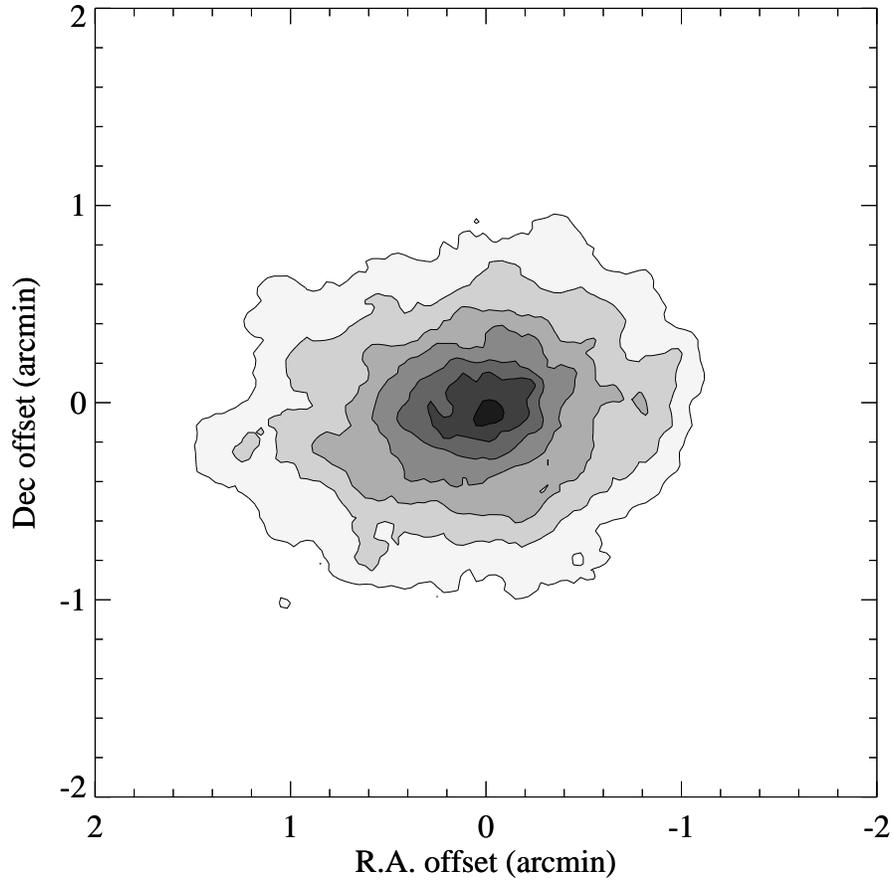

**Figure 2.** A greyscale representation of PKS0745 from the ROSAT HRI data. The pixel size is $2 \times 2$ arcsec$^2$ and the image has been adaptively smoothed to best-illustrate the cluster morphology. The surface brightness in the raw image ranges from 0 to 17 count pixel$^{-1}$. The smoothing has been adjusted to give $\geq$ 49 count pixel$^{-1}$. Contours are drawn at intervals equally separated in log space, from 0.63 to 10 count pixel$^{-1}$.

**Table 1.** Observation summary

| Instrument | Observation Date | Effective Exposure (ks) |
|---|---|---|
| ASCA SIS0 | 1993 Nov 6/7 | 24.2 |
| ASCA GIS2 | " " | 33.8 |
| ASCA GIS3 | " " | 32.5 |
| ROSAT PSPC | 1993 Oct 15 | 7.36 |
| ROSAT HRI | 1992 Oct 20 | 23.8 |
| CFHT V band | 1993 Dec 12 | 1.20 |
| CFHT I band | " " | 1.20 |
| NTT spectra | 1994 Nov 27 | 1.20 |
| " " | 1994 Nov 27 | 1.20 |
| " " | 1994 Nov 28 | 1.80 |

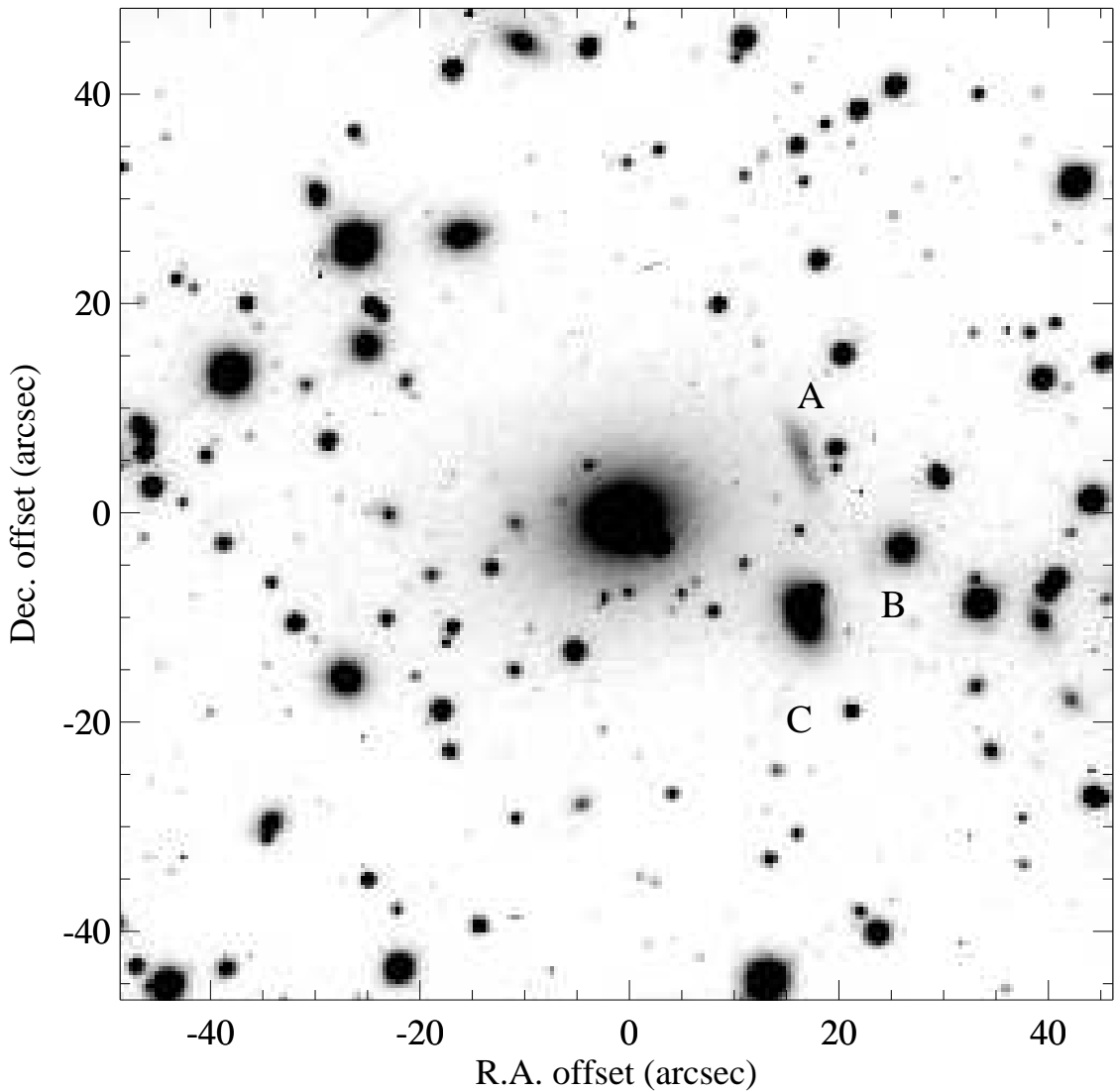

**Figure 3.** The CFHT I band image of PKS0745. The data have been binned to a pixel size of $0.412 \times 0.412$ arcsec$^2$. The CCG is the dominant galaxy in the centre of the field. The positions of the bright arc 'A' and the fainter arc candidates 'B' and 'C' are indicated.

## 3   THE MORPHOLOGY OF PKS0745

The ROSAT HRI and PSPC X-ray images of PKS0745 are shown in Figs. 1 and 2 respectively. The CFHT I band image of the cluster is presented in Fig. 3. The X-ray emission from PKS0745 is smooth, elliptical and is sharply peaked onto a position $07^h47^m31.4^s$, $-19°17'46''$ (2000.), approximately consistent with the optical CCG coordinates [$07^h47^m31.3^s$, $-19°17'40''$ (2000.)]. (Errors of $\sim 5$ arcsec in right ascension and declination are associated with the X-ray centroid, determined from the HRI data.) The sharp peak to the X-ray emission is indicative of the presence of a large cooling flow in the cluster. The X-ray emission exhibits no significant structure on large scales, such as could result from a subcluster merger event.

We have examined the morphology of the X-ray emission from the cluster, and the optical emission from the CCG, using the ELLIPSE elliptical isophote-fitting routines in IRAF. Fig. 4(a) illustrates the good agreement in the position angles of the optical and X-ray isophotes for semi-major axes (for simplicity referred to hereafter as radii) of $1 - 300$ arcsec. Note, however, the slight twisting of the optical isophotes (by $\pm 10$ degree) about a mean value of $\sim 97$ degree. The close agreement in the position angles of the CCG and cluster isophotes is similar to that observed for other relaxed, cooling-flow clusters (White *et al.* 1994; Allen *et al.* 1995). In Fig. 4(b) we plot the variation of ellipticity with radius (ellipticity is defined as $1 - b/a$ where $b$ and $a$ are the semi-minor and semi-major axes, respectively). The ellipticity of the optical CCG isophotes is well-matched to that of the X-ray emission at larger radii ($r \gtrsim 100$ arcsec) in the cluster. At smaller ($10 \lesssim r \lesssim 50$ arcsec) radii, however, the HRI observations indicate an increased ellipticity in the X-ray emission. This increase is not observable with the lower spatial resolution of the PSPC.

Although the cluster exhibits no pronounced substructure on large scales, on small scales (in the central few tens of kpc) small deviations from elliptical symmetry are observed in both the X-ray and optical data. We have investigated this structure using a simple unsharped-masking technique. The HRI X-ray image of PKS0745 (binned with $2 \times 2$ arcsec$^2$ pixels) was smoothed on two scales, using gaussian kernels of (FWHM) widths 2 and 8 pixels. An unsharped-masked X-ray image was then formed from the difference of the smoothed data sets. A similar technique was applied to the CFHT I band image (binned with $0.206 \times 0.206$ arcsec$^2$ pixels) where an unsharped-masked image was formed from the difference of the raw data and the data smoothed with a gaussian of $\sigma = 2$ pixels. The unsharped-masked X-ray and I band images are shown in Figs. 5 (a),(b).

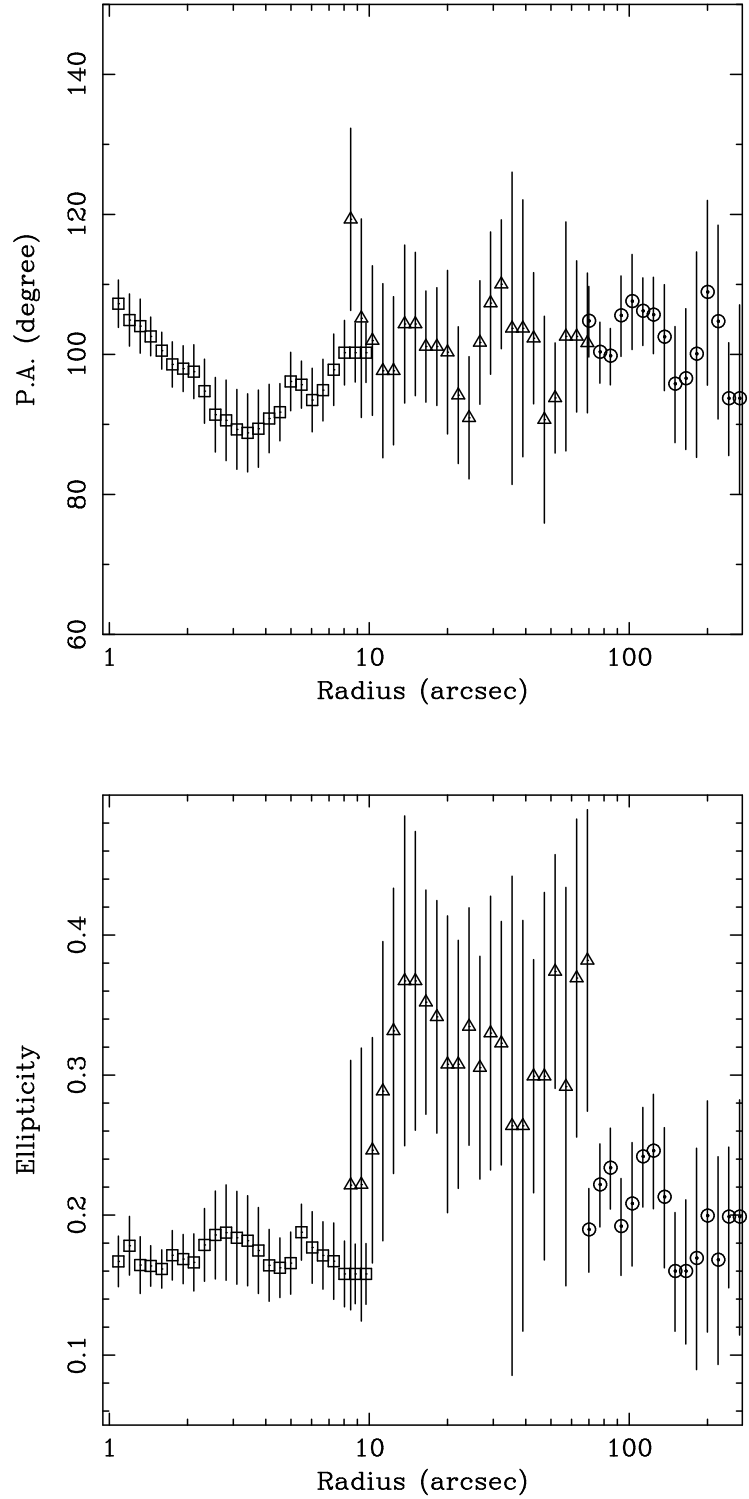

**Figure 4.** The variation of (a) position angle, (b) ellipticity with semi-major axis. Squares are the results for the optical CCG isophotes. Triangles and circles are for the HRI and PSPC X-ray data, respectively.

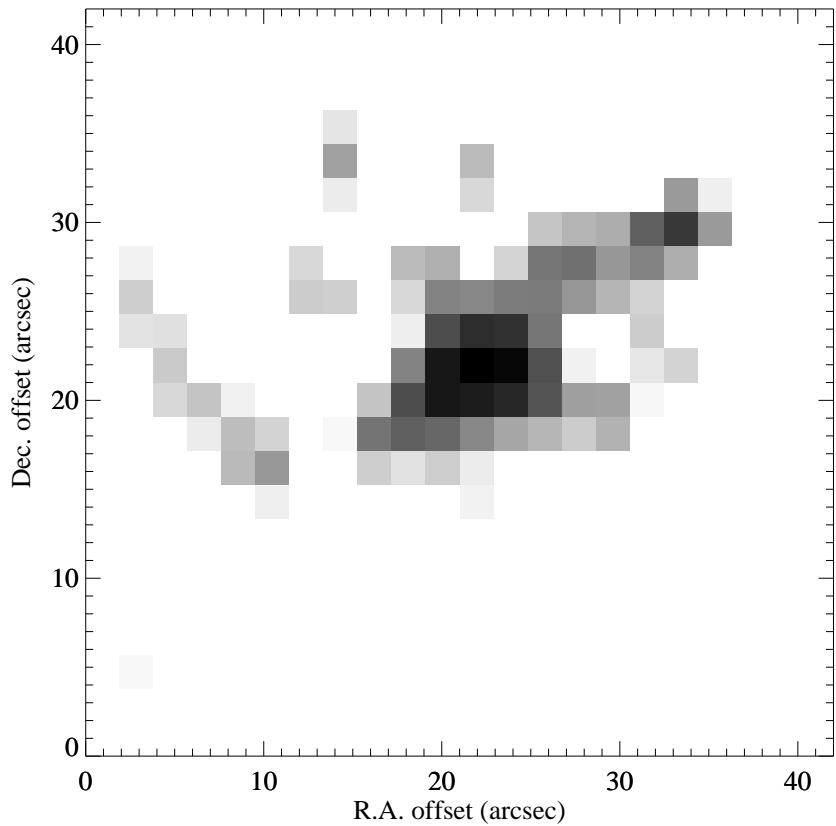

**Figure 5.** (a) The unsharped-masked HRI X-ray image of PKS0745. The image is formed from the difference of data sets smoothed with gaussian kernels of FWHM 2 and 8 pixels. (b) The unsharped-masked CFHT I-band image formed from the difference of the raw data and the data smoothed with a gaussian of width 2 pixels. Note the elongation in both images along a position angle of $120 \pm 10$ degree.

Both the X-ray and optical unsharped-masked images exhibit elongation along a position angle of $120 \pm 10$ degree (similar to the P.A. of the major axis of the cluster). In the X-ray image the enhanced emission extends $\sim 17$ arcsec to the northwest and 9 arcsec to the southeast of the nucleus. The elongation in the X-ray emission could reflect the increasing importance of angular momentum in the cooling gas at small radii (Cowie, Fabian & Nulsen 1980). However, the northwest extension may be due, in part, to a point source at a position $07^h47^m30.3^s, -19°17'39''$ (2000.). We note that this substructure accounts, in part at least, for the increased ellipticity at small radii observed in the HRI data. The I band image exhibits excess emission extending $\sim 3$ arcsec to the southeast from the nucleus. Similar structure is observed in the V band image, although at poorer statistical significance.

Romanishin (1987) showed that the CCG of PKS0745 has a strong, centrally-concentrated excess of blue emission. In Fig. 6 we plot the background-subtracted radial surface brightness profiles of the CCG in the V and I bands. (The profiles have been normalized at a radius of 20 kpc.) The data show a clear excess of blue light in the central $\sim 10$ kpc of the CCG, with $\Delta(V - I) = -0.51$ in the central (1.5 kpc) bin, in agreement with the results of Romanishin (1987; see also Fabian *et al.* 1985). Note that the blue emission is marginally brighter in the direction of the excess flux seen in the unsharped-masked images [Fig. 5(b)].

It is interesting to compare the unsharped masked images to the VLA radio maps of Baum & O'Dea (1991) and Taylor, Barton & Ge (1994). The CCG of PKS0745 contains a luminous, steep spectrum radio source (after which the cluster is named) with a remarkably complex, diffuse morphology. The radio source is also highly depolarized (Baum & O'Dea 1991; Taylor *et al.* 1994) supporting the X-ray identification of a large cooling flow in the cluster (Sections 4, 5). The unsharped-masked optical images and the 3.6 cm and 6 cm radio maps exhibit similar morphologies in the central $\sim 4$ arcsec (10 kpc) of the cluster. This correspondence may indicate a physical link between the radio and excess optical emission.

The high radio luminosity of PKS0745 suggests that it contains an active nucleus. The correspondence between the radio and optical morphologies could result from jet-induced star formation in the cooling flow or the scattering of beamed nuclear-emission into the line of sight (*e.g.* Crawford & Fabian 1993; McNamara & O'Connell 1993). Optical polarization studies should permit discrimination between these models. The optical spectrum of PKS0745 is similar to that of other CCGs in large cooling flows (although the line emission in PKS0745 is exceptionally luminous; Section 5.2). Allen (1995) discusses the formation of luminous emission-line nebulae in cooling flows from massive bursts of star formation. Within a starburst interpretation, the line luminosity of PKS0745 implies the presence of $> 10^6$ O stars in the CCG, with a large associated supernova rate.

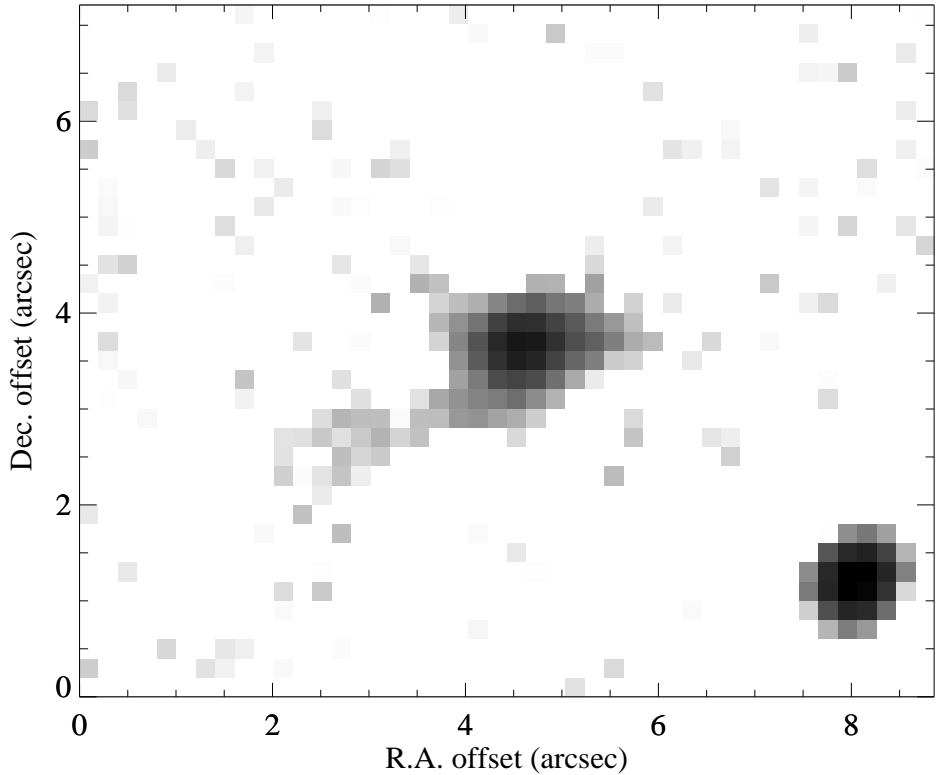

**Figure 5** - *continued.*

## 4  SPECTRAL ANALYSIS OF THE X-RAY DATA

In this section we discuss the spectral analysis of the ASCA and ROSAT X-ray data. We first describe the basic method of analysis and the modelling of the background flux in the different instruments. In Section 4.2 we present the results from a simple single-phase analysis of the data. In Section 4.3 we discuss a more detailed multi-phase analysis in which we model the spectral signature of the cooling flow in the cluster and determine the temperature profile of the ambient cluster gas.

### 4.1  The Basic Method of Analysis

The SIS observation of PKS0745 places the centroid of the X-ray emission from the cluster 4.2 arcmin to the left and 5.3 arcmin down from the top right corner of SIS0 chip 1. Cluster spectra were extracted in circular annuli of radii 0–2, 2–4 and 0–4 arcmin about the X-ray centroid. A background spectrum was extracted from a circular region of 3 arcmin radius, centred 4.2 arcmin to the right and 5.3 arcmin up from the lower left corner of chip 3. (For a detailed description of the SIS chip positions see The ABC Guide to ASCA Data Reduction).

The SIS background spectrum was modelled using a combination of gaussian, thermal, power law, and absorption components. This provides a detailed parametrization of the SIS background which can then be incorporated into the analysis of the annular cluster spectra. (Only a re-normalization of the background component, by a factor proportional to the relative areas of the source and background regions, is required.) Note that although the internal background in the SIS varies slightly from chip to chip (Gendreau 1995) the differences between SIS0 chips 1 and 3 are not significant for the analysis presented here.

The ASCA GIS detectors have an intrinsic spatial resolution which must be accounted for in addition to the resolution of the XRT. For this reason a larger annular width of 3 arcmin was adopted in the analysis of these data. The advantage of the GIS detectors is that they permit spectra to be extracted to larger radii than the SIS. At the time of writing, the calibration of the GIS instruments is well-understood within 15 arcmin of their optical axes. The position of PKS0745 in the detectors (the X-ray centroid of cluster is $\sim 7.2$ arcmin off-axis in GIS2 and 5.4 arcmin off-axis in GIS3) resulted in the extraction of spectra for annuli of $0 - 3$, $3 - 6$ and $0 - 6$ arcmin in GIS2, and $0 - 3$, $3 - 6$, $0 - 6$ and $6 - 9$ arcmin for GIS 3.

Background subtraction of the GIS data was carried out using the 'blank sky' observations (observations of high Galactic latitude fields with no bright X-ray sources) compiled during the performance verification stage of the ASCA mission. These blank sky observations provide an accurate representation of the cosmic and instrumental backgrounds in the GIS detectors between energies of 1 and 10 KeV. Note that the blank sky fields were not used in the analysis of the SIS data due to the strong variations of the soft Galactic X-ray background across the sky (which are largest near the Galactic plane) which would complicate the analysis of data in the 0.5-1 keV energy range.

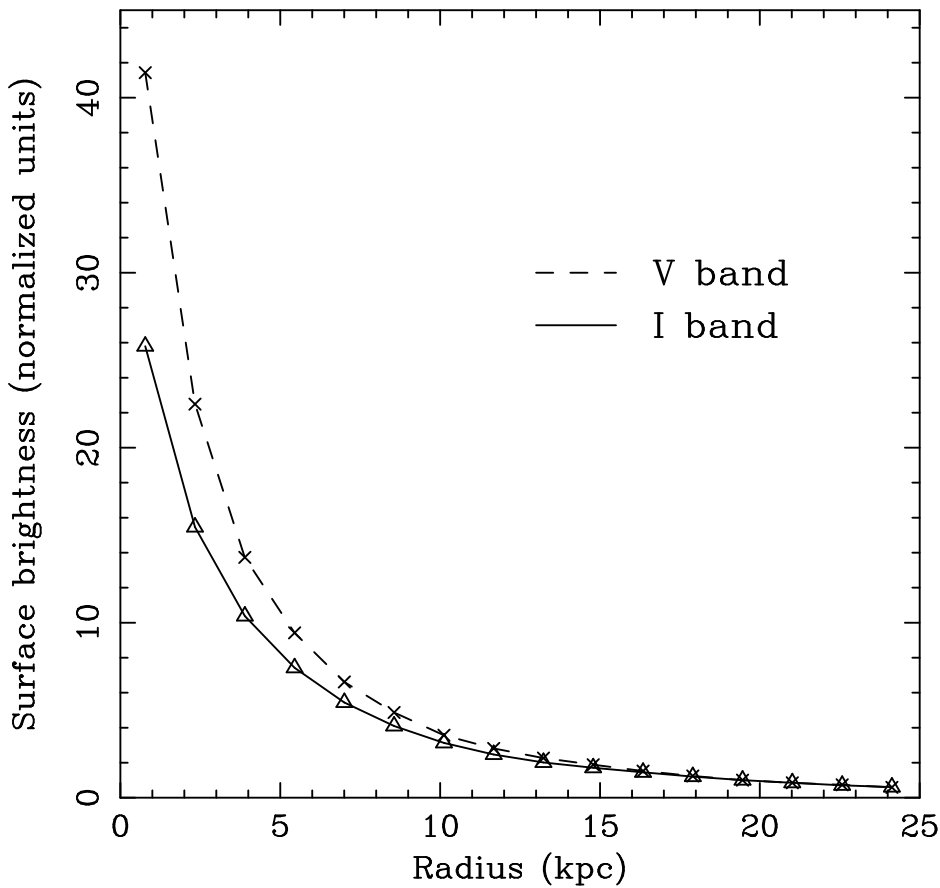

**Figure 6.** The sky-subtracted surface brightness profile of the CCG in the V and I bands. The profiles are normalized at a radius of 20 kpc. For the central (1.5 kpc) bin, $\Delta(V - I) = -0.51$.

The analysis of the PSPC spectral data was carried out using standard techniques similar to those discussed by Allen *et al.* (1993). A complete summary of the annuli used in the spectral analysis is given in Table 2.

Modelling of the X-ray spectra was carried out using the XSPEC spectral fitting package (version 8.40: Shafer *et al.* 1991). For the SIS analysis, the June 1994 version of the response matrices from GSFC were used. Only those counts in channels corresponding to energies between 0.5 and 8 keV (where the SIS is best-calibrated) have been included in the fits. For the GIS analysis, the April 1994 GSFC response matrices were used and only data in the energy range $1 - 10$ keV were analysed. For the analysis of the ROSAT data, version 36 of the PSPC response matrix was used and only data in the energy range $0.4 - 2$ keV included in the fits.

The X-ray emission from PKS0745 has been modelled using the plasma codes of Raymond & Smith (1977; with updates incorporated into XSPEC version 8.40) and Kaastra & Mewe (1993). The results for the two plasma codes are in good agreement. For clarity, only results for the Raymond & Smith (hereafter RS) code will be discussed in detail in this paper although the conclusions drawn may equally be applied to an analysis with the Kaastra & Mewe code.

Spectra were grouped before fitting to ensure a minimum of 20 counts per channel, allowing $\chi^2$ statistics to be used. We note that the use of the maximum-likelihood C-static with unbinned data sets does not significantly alter the best-fit parameter values and confidence limits obtained.

### 4.2 Single-phase analysis of the X-ray spectra

The annular spectra were first examined using a simple single-phase model. The model consists of an RS component, to account for the X-ray emission from the cluster, and a photoelectric absorption component (Morrison & McCammon 1983) to account for absorption along the line-of-sight due to our Galaxy. [The nomi-

**Table 2.** Annuli used in spectral analysis

| Annulus name | | Radius (arcmin) | Radius (kpc) |
|---|---|---|---|
| SIS0 | CORE | $0 - 2$ | $0 - 302$ |
| SIS0 | ANN1 | $2 - 4$ | $302 - 605$ |
| SIS0 | CFLOW | $0 - 4$ | $0 - 605$ |
| | | | |
| GIS2 | CORE | $0 - 3$ | $0 - 453$ |
| GIS2 | ANN1 | $3 - 6$ | $453 - 906$ |
| GIS2 | CFLOW | $0 - 6$ | $0 - 906$ |
| | | | |
| GIS3 | CORE | $0 - 3$ | $0 - 453$ |
| GIS3 | ANN1 | $3 - 6$ | $453 - 906$ |
| GIS3 | ANN2 | $6 - 9$ | $906 - 1360$ |
| GIS3 | CFLOW | $0 - 6$ | $0 - 906$ |
| | | | |
| PSPC | CORE | $0 - 1.5$ | $0 - 227$ |
| PSPC | ANN1 | $1.5 - 3$ | $227 - 453$ |
| PSPC | ANN2 | $3 - 12$ | $453 - 1814$ |
| PSPC | CFLOW | $0 - 12$ | $0 - 1814$ |

**Table 3.** Results from the Single-Phase Spectral Analysis

| SIS0 | 0–2 arcmin | 2–4 arcmin | 0–4 arcmin | |
|---|---|---|---|---|
| $kT$ | $5.57^{+0.33}_{-0.32}$ | $6.37^{+0.60}_{-0.55}$ | $5.87^{+0.41}_{-0.26}$ | — |
| $Z$ | $0.21^{+0.05}_{-0.05}$ | $0.14^{+0.06}_{-0.06}$ | $0.18^{+0.04}_{-0.04}$ | — |
| $N_{\rm H}$ | $4.34^{+0.20}_{-0.20}$ | $4.55^{+0.26}_{-0.24}$ | $4.43^{+0.16}_{-0.17}$ | — |
| $\chi^2$/DOF | 242.7/190 | 176.7/181 | 261.2/217 | — |
| | | | | |
| GIS2 | 0–3 arcmin | 3–6 arcmin | 0–6 arcmin | |
| $kT$ | $6.52^{+0.48}_{-0.46}$ | $7.36^{+1.06}_{-0.82}$ | $6.58^{+0.41}_{-0.38}$ | — |
| $Z$ | $0.26^{+0.05}_{-0.06}$ | $0.40^{+0.11}_{-0.10}$ | $0.29^{+0.08}_{-0.05}$ | — |
| $N_{\rm H}$ | $3.78^{+0.39}_{-0.39}$ | $3.07^{+0.77}_{-0.73}$ | $3.64^{+0.33}_{-0.32}$ | — |
| $\chi^2$/DOF | 372/416 | 16.3/18 | 503.1/477 | — |
| | | | | |
| GIS3 | 0–3 arcmin | 3–6 arcmin | 0–6 arcmin | 6–9 arcmin |
| $kT$ | $5.99^{+0.49}_{-0.43}$ | $7.25^{+1.04}_{-0.78}$ | $6.14^{+0.37}_{-0.37}$ | $8.93^{+4.16}_{-2.20}$ |
| $Z$ | $0.31^{+0.06}_{-0.05}$ | $0.33^{+0.10}_{-0.10}$ | $0.32^{+0.04}_{-0.05}$ | $0.19^{+0.20}_{-0.19}$ |
| $N_{\rm H}$ | $4.23^{+0.47}_{-0.39}$ | $2.99^{+0.73}_{-0.73}$ | $3.84^{+0.35}_{-0.31}$ | $3.45^{+2.62}_{-2.63}$ |
| $\chi^2$/DOF | 415/414 | 14.8/18 | 445.9/495 | 21.8/18 |
| | | | | |
| PSPC | 0–1.5 arcmin | 1.5–3 arcmin | 3–12 arcmin | |
| $kT$ | $3.9^{+2.2}_{-1.1}$ | $4.2^{+10.5}_{-1.8}$ | $> 4.1$ | — |
| $Z$ | 0.30 | 0.30 | 0.30 | — |
| $N_{\rm H}$ | $3.44^{+0.35}_{-0.34}$ | $4.42^{+0.82}_{-0.74}$ | $4.13^{+1.03}_{-0.81}$ | — |
| $\chi^2$/DOF | 15.5/22 | 14.9/24 | 24.0/24 | — |

Notes: The best-fit parameter values and 90 per cent ($\Delta \chi^2 = 2.71$) confidence limits from the single-phase spectral analysis. Temperatures ($kT$) are expressed in keV. Metallicities ($Z$) are quoted as a fraction of the Solar Value (Anders & Grevesse 1989). Column densities ($N_{\rm H}$) are in units of $10^{21}$ atom cm$^{-2}$.

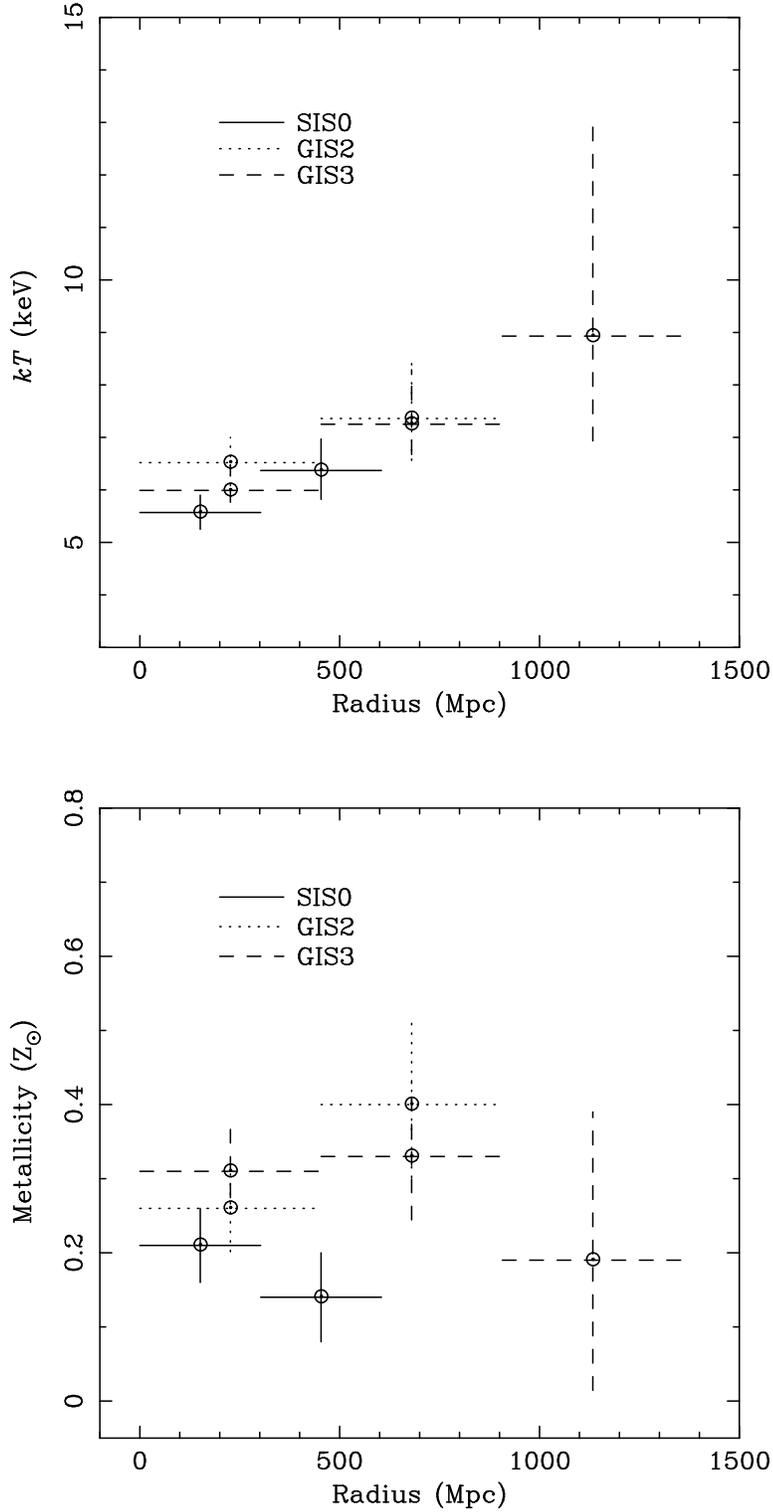

**Figure 7.** (a) The variation of temperature with radius determined from the single-phase spectral analysis. Errors are 90 per cent ($\Delta\chi^2 = 2.71$) confidence limits. (b) The variation of metallicity with radius. (c) Column density with radius.

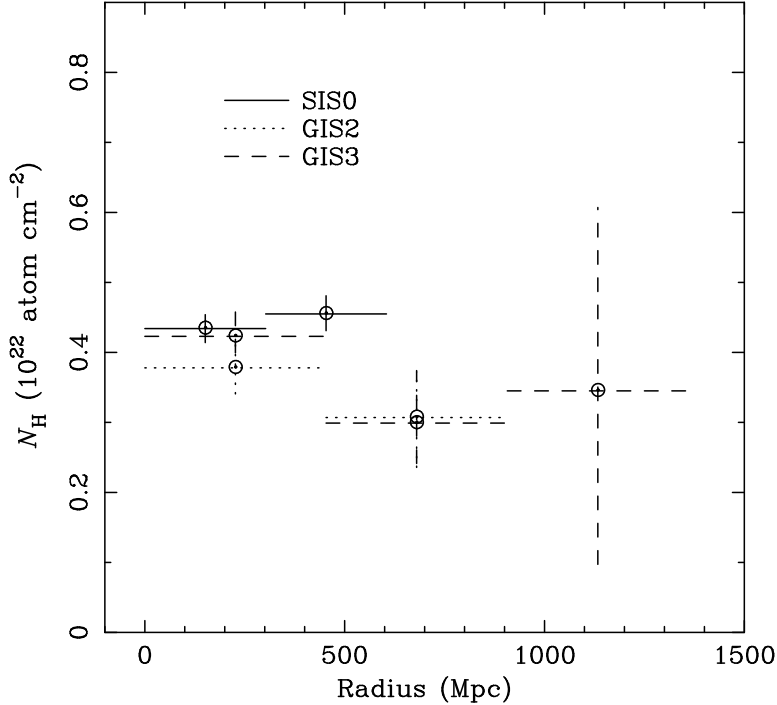

**Figure 7** - *continued.*

nal Galactic column density to PKS0745 is $4.6 \times 10^{21}$ atom cm$^{-2}$ (Stark *et al.* 1992). However, this value is interpolated from values of 4.49, 3.68, 6.01 and $4.85 \times 10^{21}$ atom cm$^{-2}$, measured at grid positions $\sim 0.5$ degree either side of PKS0745 in right ascension, and $\sim 1$ degree above and below in declination, respectively. A significant systematic uncertainty should therefore be associated with the nominal Galactic column density.] The free parameters in the single-phase model are the temperature, metallicity and emission measure of the X-ray gas, and the column density of the absorbing component (which is assumed to be at zero redshift). The redshift of the X-ray emission from the cluster was fixed at the optically-determined value for the CCG of $z = 0.1028$ (Hunstead, Murdoch & Shobbrook 1978).]

The best-fit parameter values and 90 per cent ($\Delta\chi^2 = 2.71$) confidence limits from single-phase analysis of PKS0745 are summarized in Table 3. The temperature, metallicity and column density profiles are plotted in Figs. 7(a)–(c). The results show a consistent trend for increasing temperature with radius. The GIS2 and GIS3 data are consistent with a constant metallicity of 0.3 solar at all radii, whereas the SIS0 data suggest a slightly lower metallicity of $\sim 0.2$ solar. The column densities inferred from the SIS data are in good agreement with the Galactic value, although the GIS and PSPC data measure values slightly below this mark.

**4.3   Multiphase analysis of the X-ray spectra**

Although the single-phase modelling provides a useful parametrization of the properties of the cluster gas, the results should be interpreted with caution. The deprojection analysis of Section 5 shows that PKS0745 contains a massive ($\sim 1000$ M$_\odot$ yr$^{-1}$) cooling flow. The gas in the central regions of the cluster is therefore substantially *multiphase* (*i.e.* contains a range of densities and tem-peratures). The presence of cooling, inflowing material in the inner 200–300 kpc is likely to reduce the observed emission-weighted temperature to a value below that of the ambient gas. Fortunately, the spectral capabilities of ASCA allow us to examine more detailed models for the X-ray emission from PKS0745, in which the spectral signature of the cooling gas is explicitly accounted for.

The data for the $0 - 4$ arcmin (SIS) and $0 - 6$ arcmin (GIS) annuli were analysed with two different multiphase models. (These regions contain the whole of the cooling flow and are the least-complicated by projection effects.) The first multiphase model, which provides a simple parametrization of the spread of temperatures in the cluster core, consists of two RS components. The temperatures and emission measures of the two components are independent free parameters in the fits, although the metallicities and absorption values are initially linked together. (Note that this two-component model is equivalent to introducing two extra parameters into the single-phase analysis; the temperature and emission measure of the second RS component). The results obtained with this model are detailed in Table 4 (MODEL 2RS). The inclusion of the second RS component significantly improves the fits in comparison to the single-phase results. This is indicated by the $F$-test results (Bevington 1969) listed in Table 4. [For guidance, an $F$ value of 3.0(3.8) for the inclusion of 2 (1) extra parameter(s) indicates an improvement at the 95 per cent confidence level.] The GIS data are well described by the 2RS model and the fit is not improved by the addition of further free parameters. The SIS data, however, show further improvement if the column densities on the two components are allowed to vary independently. (The significance of including this further free parameter is indicated by the F value in column 5 of Table 4). The best-fit to the SIS data then consists of two components with temperatures of $\sim 9$keV and 2 keV. The column densities on the hotter component is approximately consistent the

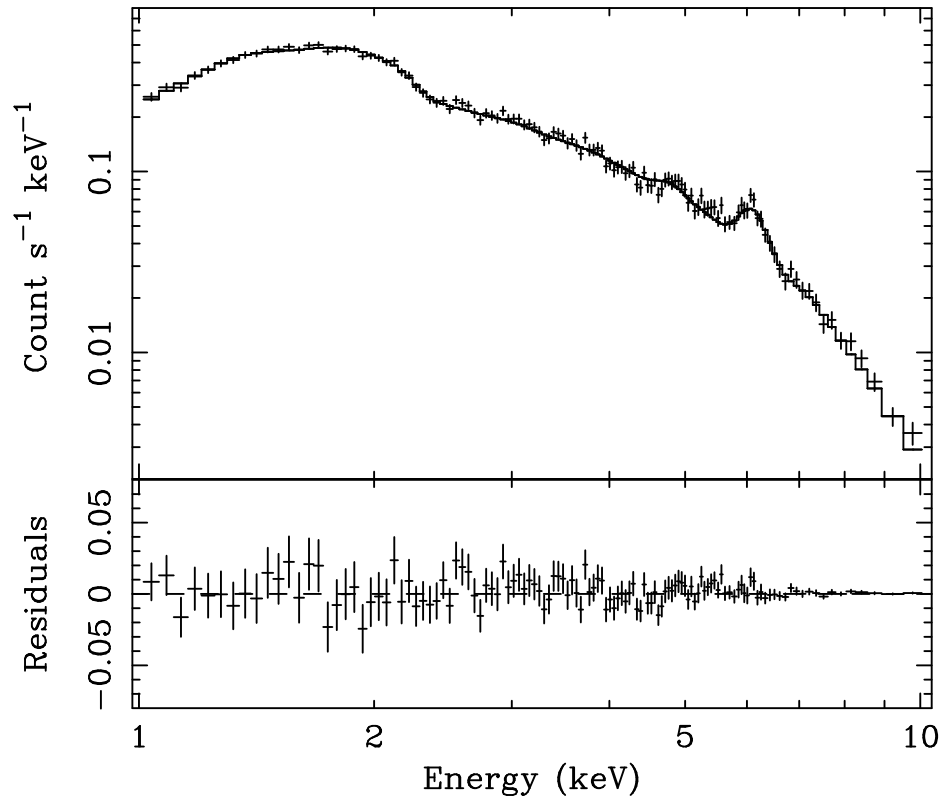

**Figure 8.** (Upper panel) The ASCA GIS spectrum of the central 6 arcmin of PKS0745 with the best fitting CFLOW model, folded through the detector response, overlaid. (Lower panel) Residuals from the best fitting model. For display purposes, the data have been binned by a factor 4 along the energy axis.

Galactic value, although the cooler component requires an excess absorption of $\sim 4 \times 10^{21}$ atom cm$^{-2}$.

The data were then analysed with a more-sophisticated model in which the spectrum of the cooling flow was accounted for explicitly. This model consists of an RS component (to model the emission from the ambient ICM in the region of interest) and a cooling-flow component (hereafter CFLOW) following Johnstone *et al.* (1992). Note that although more sophisticated than the 2RS model, the CFLOW model initially introduces only one extra free parameter into the fits relative to the single-phase analysis; the mass deposition rate of cooling gas. The upper temperature of the cooling gas, the metallicity, and the column density on the cooling flow are initially forced to the same values as the ambient emission modelled by the RS component. The results obtained with the CFLOW model, and the statistical improvement obtained for the introduction of the CFLOW component, are listed in Table 4. The fits were not statistically improved by the introduction of further free parameters. In Fig. 8 we show the GIS3 spectrum for the $0 - 6$ arcmin annulus with the best-fit CFLOW model overlaid. (Note that the PSPC spectra are also consistent with the presence of a large cooling flow, but do not provide further useful constraints on the size or properties of this component.)

It is interesting to note (particularly for the GIS data) that the column densities inferred from the multiphase models are in better agreement with the Stark *et al.* (1992) value than the single-temperature fits to the $0 - 6$ arcmin region. The CFLOW fits to the SIS data indicate that a small (relative to the Galactic value) column density of $\sim 6 \times 10^{20}$ atom cm$^{-2}$ may be associated with the central region of the cluster. Note that if the excess absorption is assumed to be intrinsic to the cooling flow (*c.f.* White *et al.* 1991; Allen *et al.* 1993; Fabian *et al.* 1994), and the column density acting on the ambient cluster emission is fixed at the Galactic value, we infer an intrinsic column density to the cooling flow of $1.1^{+0.4}_{-0.4} \times 10^{21}$ atom cm$^{-2}$. The metallicities measured with the multiphase models are slightly larger than those determined with the single temperature fits. Importantly, the mass deposition rates measured with the ASCA spectra are in good agreement with values determined independently from the deprojection analysis in Section 5.

The most important result from the multiphase analysis is the ambient cluster temperature profile determined with the CFLOW model, plotted in Fig. 9. This figure should be compared to the results from the single-phase analysis in Fig. 7(a). We see that when the effects of the cooling flow are taken into account using the CFLOW model *the cluster is observed to be approximately isothermal*. The cooling flow does not effect the results for the outer annulus of the GIS3 data. Combining the data from the SIS0, GIS2 and GIS3 detectors, we determine an ambient cluster temperature in the region of the cooling flow of $kT = 10^{+4.5}_{-2.5}$ keV. For comparison, the analysis of observations made with the EXOSAT Medium Energy detector (Arnaud *et al.* 1987) determined a mean temperature for the whole cluster of $8.6^{+1.1}_{-0.9}$ keV.

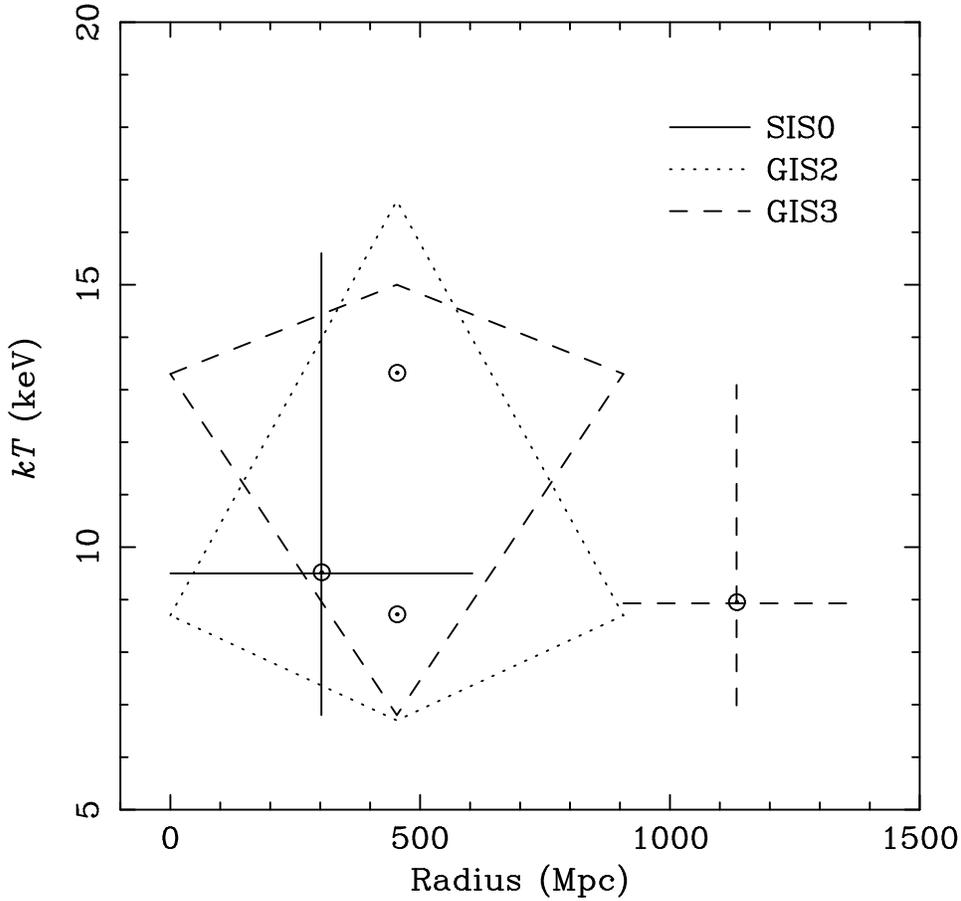

**Figure 9.** The temperature profile of the ambient cluster gas determined from the multiphase spectral analysis. Errors are 90 per cent ($\Delta\chi^2 = 2.71$) confidence limits. The solid bars are for the SIS0 data. The dotted bars and dashed bars are for the GIS2 and GIS3, respectively.

## 5 DEPROJECTION ANALYSIS OF THE X-RAY DATA

In this Section the constraints on the ambient cluster temperature profile (Fig. 9) are used in combination with a deprojection analysis of the X-ray images to determine the distribution of mass in PKS0745. The deprojection technique is used to examine the properties of the cooling flow in the cluster.

### 5.1 X-ray constraints on the cluster mass profile

The deprojection technique of Fabian *et al.* (1981) is a standard method for the analysis of X-ray images of clusters of galaxies. In previous work we have used the deprojection method to measure the properties of the ICM (density, pressure, temperature) given assumptions of spherical symmetry, hydrostatic equilibrium in the cluster gas and a specified form for the gravitational potential (constrained by the velocity dispersion and/or mean temperature) of the cluster. However, if the temperature profile of the cluster is known (from spatially-resolved spectroscopy) the deprojection technique can be inverted to measure the distribution of mass in the cluster.

We have examined the distribution of mass in PKS0745 using the constraints on the ambient cluster temperature profile from Section 4.3. We find that the mass distribution can be parametrized in a number of simple ways. A good parametrization (and one that can be intuitively understood) is obtained with a 2-component model consisting of an isothermal sphere (Binney & Tremaine 1987) with a velocity dispersion of $1100 \pm 200\,\mathrm{km\,s^{-1}}$ and a core radius of $130 \pm 20$ kpc (modelling the bulk of the mass of the cluster) and a smaller, singular isothermal component, contributing $\sim 2 \times 10^{11}$ $\mathrm{M_\odot\,kpc^{-1}}$ and truncated at an outer radius of 100 kpc (modelling the mass contribution of the large CCG). The total mass within 130 kpc is $8.0 \times 10^{13}$ $\mathrm{M_\odot}$. Both mass components are centred on the X-ray centroid.

Alternatively, the mass distribution can be approximated using a simple analytical form

$$\rho(r) = C \left[ \frac{1}{1 + (r/r_c)^2} \right] \qquad (2)$$

where $\rho(r)$ is the mass density in units of $\mathrm{M_\odot\,kpc^{-3}}$, $r$ and $r_c$

**Table 4.** Results from the Multiphase Spectral Analysis

| 2RS | GIS2 | GIS3 | SIS0 | SIS0 ($N_H$ indep.) |
|---|---|---|---|---|
| $kT_1$ | $7.0^{+0.7}_{-0.7}$ | $> 13.5$ | $8.1^{+\infty}_{-1.6}$ | $8.6^{+\infty}_{-1.6}$ |
| $kT_2$ | $0.83^{+0.30}_{-0.24}$ | $3.9^{+0.8}_{-0.7}$ | $2.0^{+2.4}_{-0.7}$ | $1.6^{+1.8}_{-0.5}$ |
| $Z$ | $0.29^{+0.06}_{-0.06}$ | $0.44^{+0.11}_{-0.10}$ | $0.20^{+0.06}_{-0.05}$ | $0.22^{+0.05}_{-0.06}$ |
| $N_{H,1}$ | $5.85^{+1.62}_{-1.34}$ | $4.63^{+0.49}_{-0.51}$ | $4.70^{+0.25}_{-0.22}$ | $3.92^{+0.43}_{-1.40}$ |
| $N_{H,2}$ | — | — | — | $8.71^{+3.19}_{-2.71}$ |
| $\chi^2$/DOF | 495.6/475 | 433.5/493 | 248.8/215 | 243.8/214 |
| $F$ | 3.6 | 7.1 | 5.4 | 4.4 |

| MODEL CFLOW | GIS2 | GIS3 | SIS0 | SIS0 ($N_H$ intrin.) |
|---|---|---|---|---|
| $kT$ | $8.7^{+7.9}_{-2.0}$ | $13.3^{+1.7}_{-6.5}$ | $9.5^{+6.1}_{-2.7}$ | $9.5^{+5.9}_{-2.5}$ |
| $Z$ | $0.36^{+0.13}_{-0.08}$ | $0.47^{+0.06}_{-0.14}$ | $0.26^{+0.07}_{-0.07}$ | $0.26^{+0.06}_{-0.07}$ |
| $N_H$ | $4.51^{+0.69}_{-0.78}$ | $5.15^{+0.36}_{-0.84}$ | $5.21^{+0.28}_{-0.41}$ | 4.6 |
| $\Delta N_H$ | — | — | — | $1.06^{+0.39}_{-0.39}$ |
| $\dot{M}$ | $1130^{+580}_{-930}$ | $1710^{+160}_{-1030}$ | $1370^{+280}_{-720}$ | $1350^{+250}_{-580}$ |
| $\chi^2$/DOF | 499.3/476 | 439.0/494 | 253.0/216 | 251.0/216 |
| $F$ | 3.6 | 7.8 | 7.0 | |

Notes: The best-fit parameter values and 90 per cent ($\Delta \chi^2 = 2.71$) confidence limits from the multiphase spectral analysis. Temperatures ($kT$) are expressed in keV. Metallicities ($Z$) are quoted as a fraction of the Solar Value (Anders & Grevesse 1989). Column densities ($N_H$) are in units of $10^{21}$ atom cm$^{-2}$. Mass deposition rates ($\dot{M}$) are in $M_\odot$ yr$^{-1}$. $F$ values [following the F-test prescription of Bevington (1969)] measure the significance of the improvements to the fits (relative to the single-phase results) obtained with the multiphase models. For MODEL 2RS (columns 2–4) $F$ values measure the significance of including the second RS component (2 extra free parameters). In column 5, the $F$ value measures the further improvement obtained with the SIS0 data when the column densities on the two RS components are allowed to vary independently (1 further free parameter). For MODEL CFLOW, $F$ values describe the significance of including the CFLOW component (1 extra free parameter) in the fits. [For guidance, an $F$ value of 3.0(3.8) for the inclusion of 2 (1) extra parameter(s) indicates an improvement at the 95 per cent confidence level.] Column 5 for MODEL CFLOW describes the results for a model in which the column density on the ambient cluster gas is fixed at the Galactic value ($4.6 \times 10^{21}$ atom cm$^{-2}$) and excess absorption ($\Delta N_H$; at the redshift of the cluster) is associated with the cooling flow as a free parameter.

are the radius and core radius in kpc with $r_c = 25^{+15}_{-15}$ kpc, and $C$ is a normalization factor of value $10^{+5}_{-3} \times 10^7$ M$_\odot$ kpc$^{-3}$.

The radial profiles of the total gravitating mass, and the X-ray emitting gas mass, in PKS0745 are plotted in Fig. 10(a). It is useful to compare the results on the mass profile, using the temperature constraints from the multiphase analysis, to the results that would have been obtained if the temperature profile from the single-phase analysis were (wrongly) used instead. An ambient temperature profile matching the single-phase results is obtained using an isothermal mass distribution, with a core radius of $\sim 250$ kpc and no additional mass component associated with the CCG. The velocity dispersion (which principally controls the temperature at larger radii) is unchanged. Alternatively, using the analytical parametrization of Equation 2, the single-phase temperature profile can be recovered for $r_e \sim 100$ kpc (and the same value of $C$). *The multiphase spectral analysis therefore implies a substantially more centrally-concentrated mass distribution in PKS0745 than would be inferred from a simple single-phase analysis.* Note, with reference to Section 6, that the projected mass within the critical radius for gravitational lensing (45.9 kpc) is *reduced by a factor* $\sim 3$ by using the single-phase, rather than the multiphase temperature constraints.

In Fig. 10(b) we plot the gas mass/total mass ratio in PKS0745 as a function of radius. The ratio of $0.18^{+0.06}_{-0.05}$ (rising to $0.23^{+0.07}_{-0.06}$ at 2 Mpc) illustrates an overdensity of baryons, with respect to the predictions from standard models of primordial nu-cleosynthesis, similar to other X-ray luminous clusters (White & Frenk 1991; White *et al.* 1993; White & Fabian 1995). The result for PKS0745 is important given the firm constraints on the distribution of total mass in the cluster presented in this paper.

## 5.2 The cooling flow in PKS0745

Previous studies of PKS0745 with the Einstein Observatory and EXOSAT (Fabian *et al.* 1985; Arnaud *et al.* 1987) have shown that the cluster contains one of the largest known cooling flows. We have used the deprojection technique to investigate, in detail, the properties of the cooling flow in PKS0745. The results, summarized in Table 5, show that the cluster contains a cooling flow of $\sim 1000$ M$_\odot$ yr$^{-1}$ (in excellent agreement with the results from spectral analysis in Section 4.3) and that the mass deposition is concentrated within the inner $\sim 200$ kpc of the cluster.

The azimuthally averaged X-ray surface brightness profile (background-subtracted and corrected for telescope vignetting) and the principal results from the deprojection analysis are summarized in Figs 11(a)–(h). Note that X-ray emission from the cluster is observed to a radius of $> 0.3$ degree in the PSPC image, but is not detected beyond 0.4 degree. (Between 0.3 and 0.4 degree the PSPC is masked by its rib-support structure.) For convenience we assume a maximum cluster extent of 3 Mpc, corresponding to 0.33 degree

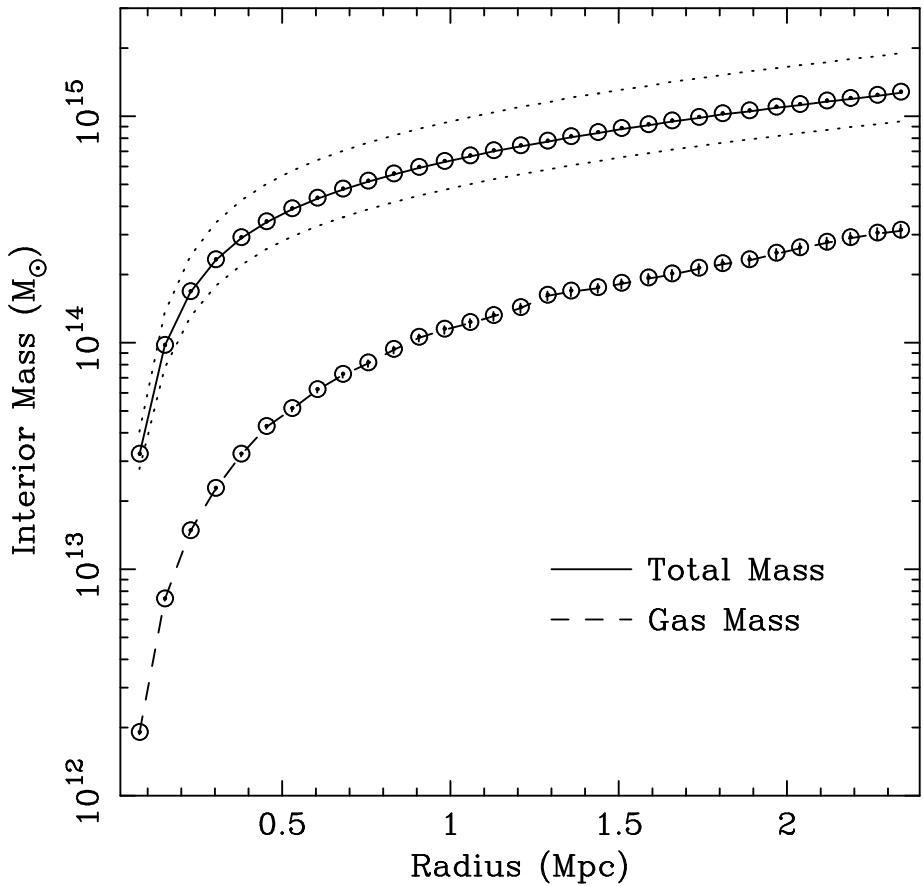

**Figure 10.** The total mass and gas mass distributions for PKS0745 determined from the X-ray data. The solid lines mark the values for a 10 keV cluster. The dotted lines trace the total mass profiles for clusters of 8 keV (lower) and 14 keV (upper) respectively. (b) The gas mass/total mass ratio as a function of radius.

at $z = 0.1028$. The cluster is assumed to remain isothermal to its outer radius.

The high mass deposition rate in PKS0745 is supported by the large H$\beta\lambda$4861 flux associated with the CCG ($F_{H\beta} \sim 2.5 \pm 1.0 \times 10^{-14}$ erg cm$^{-2}$ s$^{-1}$; Fabian *et al.* 1985). This implies an extinction-corrected H$\beta\lambda$4861 luminosity of $2 \pm 1 \times 10^{43}$ erg s$^{-1}$. A similarly high line extinction-corrected luminosity is inferred from the observations of Heckman *et al.* (1989). These observations identify PKS0745 as the most line-luminous central cluster galaxy known. [The correction for Galactic extinction follows the relationship between equivalent hydrogen column density and $E(B - V)$ from Bohlin, Savage & Drake (1978) and the relationship between $E(B - V)$ and $A_\lambda$ from Zombeck (1990).]

## 6    GRAVITATIONAL LENSING IN PKS0745

In this section we present the evidence for gravitational lensing in PKS0745. The properties of the brightest lensed source in the field are discussed in detail. We use both simple spherically-symmetric, and more sophisticated elliptical lensing models to measure the projected mass of the cluster within the radius defined by the bright arc. These results are compared to the mass measurements from the X-ray data.

### 6.1    The imaging data

The CFHT optical I band image of PKS0745 (Fig. 3) shows the presence of a bright, tangentially-oriented, arc-like feature at a radius of 18.2 arcsec (45.9 kpc) and a position angle of 290 degree about the central galaxy. The feature is also visible in the V band although at substantially lower signal-to-noise (due in part to the large extinction along the line of sight by the Galaxy; $A_V \sim 3$). The arc is clearly resolved by the CFHT data with a length, $l \sim 15.5$ arcsec and a width, $w \sim 3.5$ arcsec. (The length and width are defined at the zero-intensity limit where the arc blends into the sky). A gaussian fit to a cross section across the arc (at a position angle of 110 degrees) measures a FWHM of $1.7 \pm 0.1$ arcsec. Two other (substantially fainter) arc candidates are observed in the I band image at radii of 24.1 arcsec (60.7 kpc) and 23.6 arcsec (59.5 kpc), and position angles of 261 degree and 219 degree, respectively. [We note here that S. Baum first verbally reported the possible existence of a gravitational arc in PKS0745 during a discussion of emission-line nebulae in clusters of galaxies; Baum (1992).]

The bright arc exhibits a complex morphology, with clear colour variation along its length. In Figs. 12 (a),(b) we plot the surface brightness of the arc in the I and V bands for an aperture of width 1.5 arcsec positioned along the arc (P.A.= 20 degree; the same slit width and P.A. as used in the spectral observations discussed be-

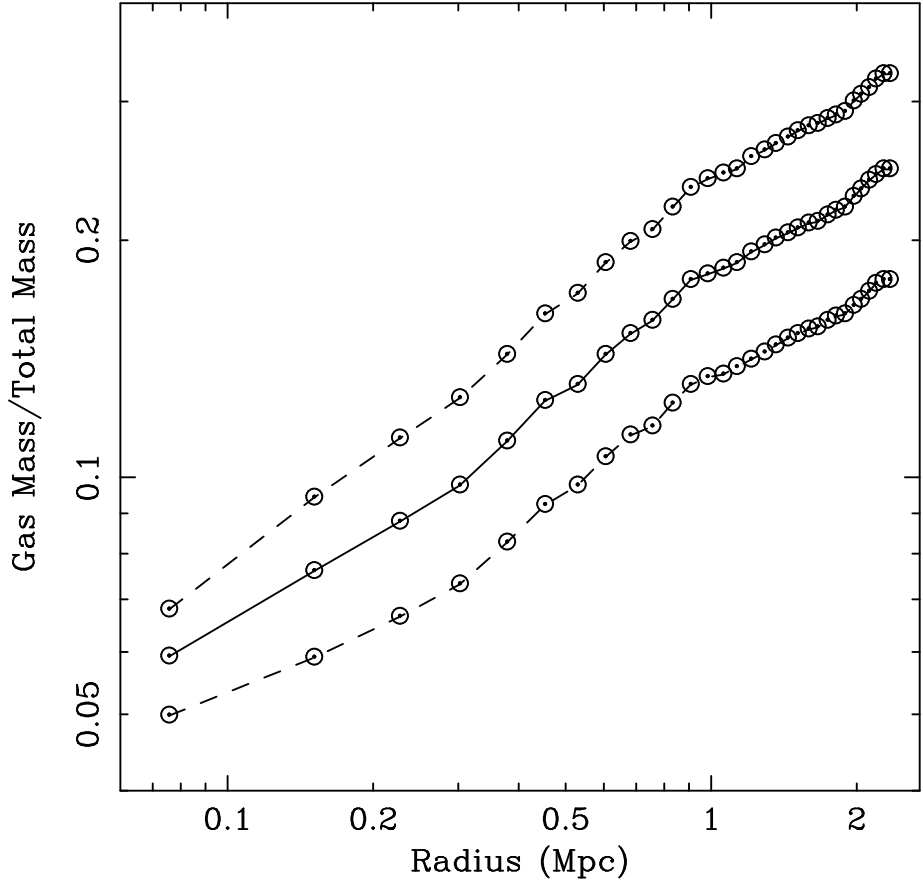

**Figure 10** - *continued.*

**Table 5.** Cooling flow results

| Instrument | Cosmology | | Cooling Flow Parameters | | |
|------------|-----------|------|------|------|------|
| | $H_0$ | $q_0$ | $r_{cool}$ kpc | $t_{cool}$ $10^9$ yr | $\dot{M}$ $M_\odot$ yr$^{-1}$ |
| PSPC | 50 | 0.5 | $212^{+52}_{-23}$ | $1.95^{+0.07}_{-0.06}$ | $995^{+113}_{-70}$ |
| PSPC | 50 | 0 | $271^{+78}_{-77}$ | $1.99^{+0.07}_{-0.07}$ | $1115^{+257}_{-133}$ |
| PSPC | 100 | 0.5 | $86^{+8}_{-29}$ | $1.43^{+0.05}_{-0.05}$ | $229^{+18}_{-32}$ |
| PSPC | 100 | 0 | $109^{+26}_{-12}$ | $1.45^{+0.04}_{-0.06}$ | $255^{+27}_{-17}$ |
| HRI | 50 | 0.5 | $180^{+11}_{-9}$ | $1.11^{+0.18}_{-0.14}$ | $727^{+116}_{-204}$ |
| HRI | 50 | 0 | $254^{+110}_{-78}$ | $1.13^{+0.26}_{-0.16}$ | $926^{+780}_{-308}$ |
| HRI | 100 | 0.5 | $87^{+9}_{-21}$ | $0.86^{+0.15}_{-0.12}$ | $196^{+25}_{-36}$ |
| HRI | 100 | 0 | $95^{+45}_{-7}$ | $0.86^{+0.17}_{-0.11}$ | $193^{+150}_{-40}$ |

Notes: A summary of the results on the cooling flow in PKS0745 from the deprojection analysis. All results are for an assumed ambient cluster temperature of 10 keV. Cooling radii ($r_{cool}$) are in kpc. Cooling times ($t_{cool}$) for the central bin (15 arcsec for the PSPC data and 8 arcsec for the HRI are in $10^9$ yr. Mass deposition rates ($\dot{M}$) in $M_\odot$ yr$^{-1}$ . Errors on the cooling times are the 10 and 90 percentile values from 100 Monte Carlo simulations. The upper and lower confidence limits on the cooling radii are the points where the 10 and 90 percentiles exceed, and become less than, the Hubble time, respectively. Errors on the mass deposition rates are the 90 and 10 percentile values at the upper and lower limits for the cooling radius. A Galactic column density of $4.6 \times 10^{21}$ atom cm$^{-2}$ (Stark *et al.* 1992) has been assumed.

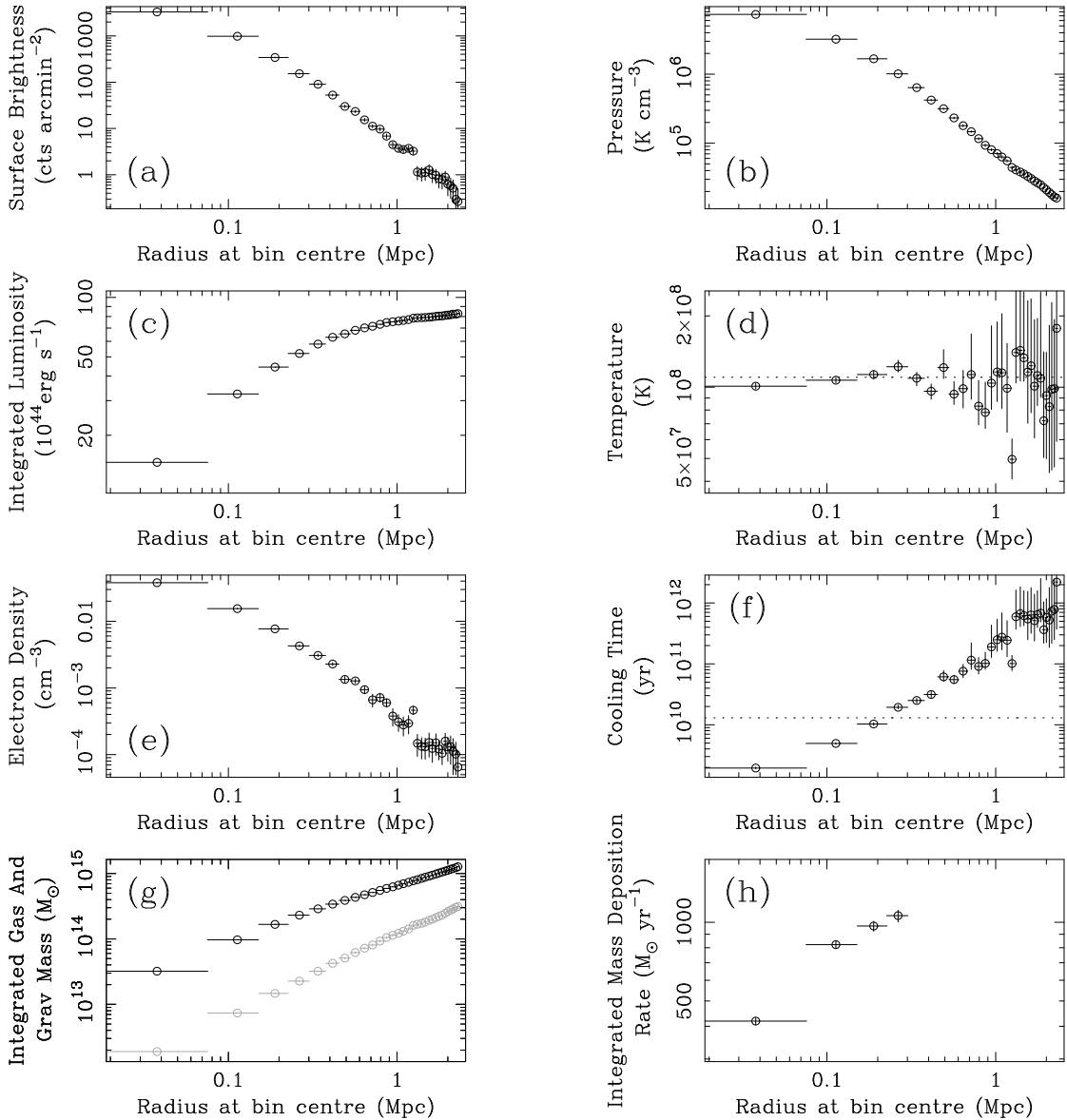

**Figure 11.** A summary of the results from the deprojection analysis of the PSPC data. The cluster potential is modelled with the (10 keV) mass profile presented in Fig. 10. From left to right, top to bottom, we plot; (a) surface brightness, (b) pressure, (c) integrated luminosity, (d) temperature, (e) electron density, (f) cooling time, (g) integrated gas and gravitational mass and (h) integrated mass deposition rate. Data points are mean values and $1\sigma$ errors (in each radial bin) from 100 Monte Carlo simulations, except for (d), (f) and (h) where the median and 10 and 90 percentile values have been plotted.

low). We measure apparent magnitudes for the arc of $I = 17.7 \pm 0.2$ and $V = 21.1 \pm 0.2$, and determine an intrinsic (V-I) colour (corrected for Galactic extinction; $A_V \sim 3.1$, $A_I \sim 1.8$) of $\sim 2.1$. Note that errors on the apparent magnitudes are dominated by uncertainties in the flux calibration. The colour variation along the arc and its extension suggest that the lensed source is a moderately lensed galaxy rather than a highly magnified compact source. The (V-I) colour is consistent with an $S_{ab}$ galaxy at redshift $z \sim 0.4$ (Lilly 1993). No other objects with the same color and surface brightness are visible in the central region, suggesting that the arc is a 'single' arc. The properties of the arc, and the fainter arc candidates, are summarized in Table 6.

### 6.2 Spectroscopy of the bright arc

A series of spectra of the bright arc were obtained with the NTT over the period 1994 November 27–28. The co-added, calibrated data are presented in Fig. 13(a). Two emission lines are observed in the spectrum, at wavelengths of 5344A and 6962A, which we identify as [OII]$\lambda$3727 and H$\beta\lambda$4861 emission at a redshift of $0.433 \pm 0.001$.

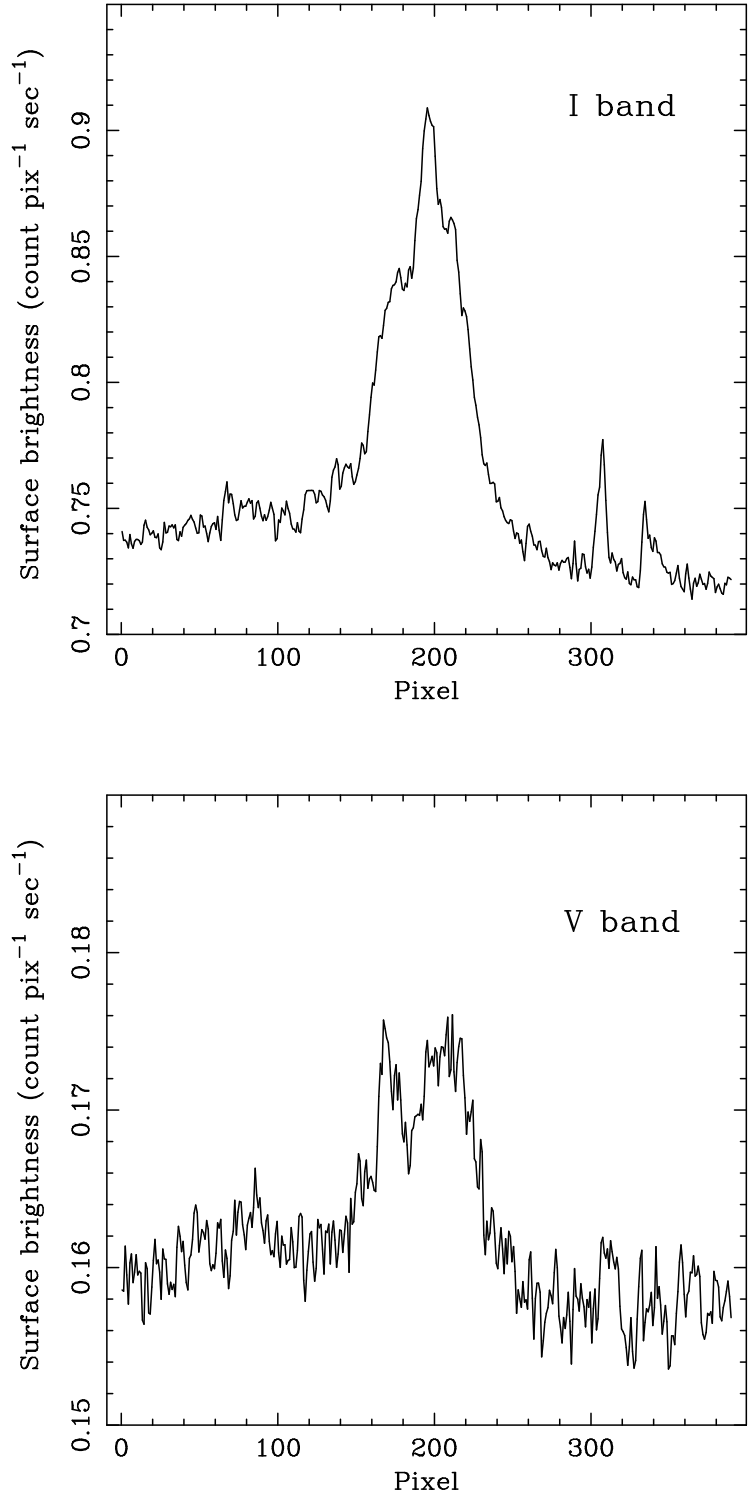

**Figure 12.** (a) The I band surface brightness of the bright arc 'A' along a slice of width 1.5 arcsec, at a P.A. of 20 degree. (b) The surface brightness in the V band for the same region. The raw (debiased) data (pixel size is $0.103 \times 0.103$ arcsec$^2$) have been used. Increasing pixel number denotes a northeasterly direction.

**Table 6.** The observed properties of the arc candidates

| Arc 'A' | | |
| --- | --- | --- |
| Quantity | Unit | Value |
| Radius from CCG, $r_{proj}$ | arcsec (kpc) | $18.2 \pm 0.1$ (45.9) |
| Position Angle from CCG | degree | $290 \pm 1$ |
| length, $l$ (maximum extent) | arcsec | $15.5 \pm 0.4$ |
| width, $w$ (maximum extent) | arcsec | $3.5 \pm 0.2$ |
| orientation | degree | $19 \pm 1$ |
| Redshift, $z_{arc}$ | | $0.433 \pm 0.001$ |
| Apparent Magnitude (I) | Mag | $17.7 \pm 0.2$ |
| Apparent Magnitude (V) | Mag | $21.1 \pm 0.2$ |
| Colour (V-I) | Mag | $3.4 \pm 0.3$ |
| Corrected Colour (V-I) | Mag | $2.1 \pm 0.3$ |

| Arc 'B' | | |
| --- | --- | --- |
| Quantity | Unit | Value |
| Radius from CCG | arcsec (kpc) | $24.1 \pm 0.1$ (60.7) |
| Position Angle from CCG | degree | $261 \pm 1$ |
| orientation | degree | $15 \pm 3$ |

| Arc 'C' | | |
| --- | --- | --- |
| Quantity | Unit | Value |
| Radius from CCG | arcsec (kpc) | $23.6 \pm 0.1$ (59.5) |
| Position Angle from CCG | degree | $219 \pm 1$ |
| orientation | degree | $138 \pm 5$ |

Notes: A summary of the observed properties of the bright gravitational arc 'A' and the two fainter arc candidates 'B' and 'C'. The errors in magnitudes and colours for 'A' are dominated by uncertainties in the flux calibration. The corrected colour is the observed (V-I) colour corrected for Galactic extinction [$A_V = 3.1$, $A_I = 1.8$; derived from the equivalent hydrogen column density to the cluster (Stark *et al.* 1992), the relationship between equivalent hydrogen column density and $E(B - V)$ of Bohlin, Savage & Drake (1978) and the relationship between $E(B - V)$ and $A_\lambda$ from Zombeck (1990)].

The observed line widths of $12.9^{+1.3}_{-1.5}$Å and $11.7^{+1.0}_{-1.0}$Å respectively, imply an intrinsic velocity width at the source of $\sim 300$ km s$^{-1}$. An [OII]$\lambda$3727/H$\beta\lambda$4861 flux ratio of $1.5 \pm 0.2$ is observed. The continuum and line emission from the arc are consistent with a spiral galaxy with ongoing star formation.

The redshift and angular size of the arc imply a physical size, in the absence of lensing, of $\sim 100$ kpc. This far exceeds the visible size of any normal galaxy and confirms that the source is gravitationally lensed. In Fig. 13(b) we present the 2-dimensional spectrum of the arc in the region of the redshifted [OII]$\lambda$3727 line (a region free from strong sky lines.) The data provide clear evidence for velocity shear of $390 \pm 40$ km s$^{-1}$ across the arc, which is consistent with the 'single' image assumption. Velocity gradients along arcs have previously been observed for the straight arc in Abell 2390 ($z = 0.913$; Pello *et al.* 1991) and in the double arc in Cl2236 ($z = 1.116$; Melnick et al. 1993). Narashima & Chitre (1993) noted that such gradients can be used to a constrain the modelling of arcs. This constraint was used by Kneib *et al.* (1994) in their modelling of the Cl2236 arc, where it was shown that both the morphology of the arc and the velocity shear are consistent with a galaxy merger.

### 6.3 The lensing mass

The observations of gravitational lensing in PKS0745 can be used to determine the projected mass of the central part of the cluster.

For a circular mass distribution the projected mass within the tangential critical radius is given by

$$M_{proj}(r_{ct}) = \frac{c^2}{4G} \left( \frac{D_{arc}}{D_{clus} D_{arc-clus}} \right) r_{ct}^2 \qquad (3)$$

where $r_{ct}$ is the tangential critical radius (Einstein radius) and $D_{clus}$, $D_{arc}$ and $D_{arc-clus}$ are respectively the angular diameter distances from the observer to the cluster, the observer to the lensed object, and the cluster to the lensed object. For $r_{ct} \sim r_{arc} = 45.9$ kpc, we measure a projected mass of $M_{proj}(r_{ct}) = 3.0 \times 10^{13}$ $M_\odot$. Within the circular aperture defined by the critical radius we measure absolute magnitudes (corrected for Galactic extinction) of $M_V = -24.8$ and $M_I = -26.9$, implying mass-to-light ratios of $M/L_V = 42 \pm 8$ and $M/L_I = 13 \pm 3$. Note, however, that given the low Galactic latitude of PKS0745, the correction for extinction is somewhat uncertain.

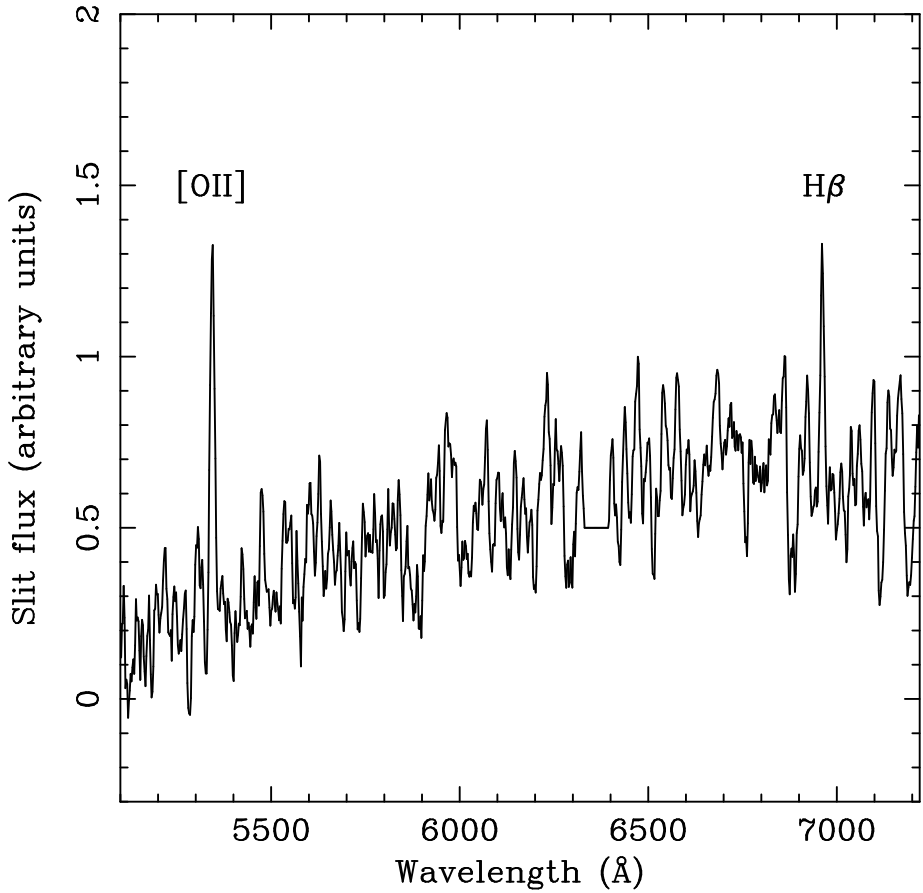

**Figure 13.** (a) The NTT spectrum of arc 'A'. The exposure time is 4200s. The spectrum has been sky-subtracted, cleaned of obvious cosmic rays, and lightly smoothed. (b) The 2-dimensional spectrum of the arc in the region of the redshifted [OII]$\lambda3727$ line. A velocity shear of $\sim 400$ km s$^{-1}$ across the arc is evident. The data have been smoothed with a median filter of width 3 pixels.

Detailed studies of systems with multiple arcs (Mellier *et al.* 1993; Kneib *et al.* 1993, 1995) have shown that circular mass distributions can not fully explain the observed arc configurations. These studies have shown that the geometry of the total mass distribution (orientation, ellipticity) approximately follows the geometry of the CCG halo and the X-ray emission from the cluster. Furthermore, as noted by Bartelmann (1995), circular mass models tend to give larger lensing masses than elliptical models. This is principally due to the shear intensity at a position along the major axis being larger than the shear intensity at the same distance along the minor axis. Hence we are biased to observe arcs along the axis of elongation of the cluster, and applying a circular mass model to these arcs can lead to overestimates of the cluster mass. (Similarly, applying the circular formula to arcs positioned along the minor axis of the mass distribution, can lead to underestimates of the lensing mass.)

Furthermore, the bright arc in PKS0745 is not as dramatically distorted as other famous cluster arcs. The color variation and velocity gradient are both consistent with a 'single' image arc. Therefore the critical line will lie at a smaller radius than the arc, which again lowers the mass estimate from the value obtained with the circular model. As we have only one arc with a measured redshift, and have no multiple images, it is not possible to rigorously determine the distribution of mass in PKS0745. However, a relatively strong constraint on the mass within the arc radius can be derived.

In Table 7 we present three different elliptical mass-models which reproduce the observed arc parameters. We assume that the center of the total mass distribution is given by the center of the CCG, and the orientation of the mass by the mean orientation of the CCG and X-ray isophotes. The analytical mass model used is the pseudo isothermal elliptical distribution (PIEMD) of Kassiola & Kovner (1993) which corresponds to the elliptical form of the 2D projection of Equation 2. Model 1 has a small core (25kpc) and a small ellipticity $(1 - b/a = 0.17)$; Model 2 has a larger core (40kpc) and a small ellipticity; Model 3 has a large core and large ellipticity $(1 - b/a = 0.3)$. The mass estimates and amplification factors for the arc, summarized in Table 7, are consistent with a total mass within the arc of $M_{proj}(r_{arc}) \sim 2.5 \times 10^{13}$ $M_\odot$. There is little scatter in the results for the different models. The extinction-corrected magnitude of the arc in the absence of lensing, $I \sim 18.3 - 20.0$, is consistent with an $S_{ab}$ galaxy at redshift $z \sim 0.4$ (K. Glazebrook, private communication).

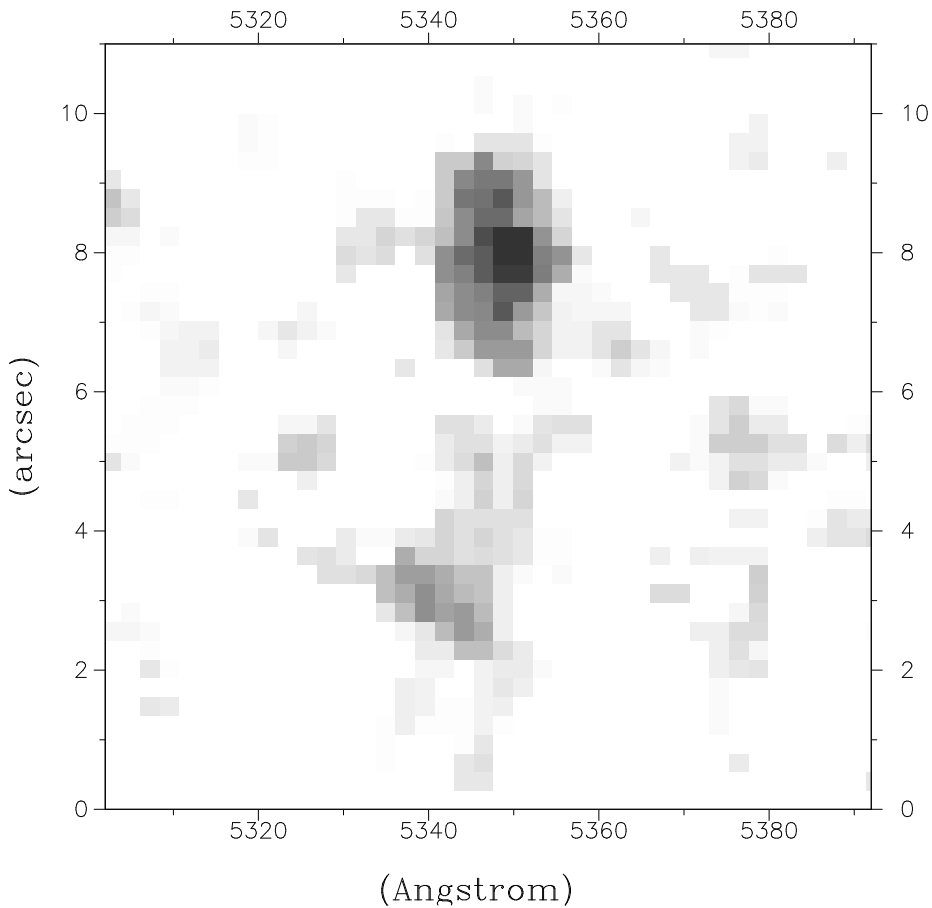

**Figure 13** - *continued.*

### 6.4    The comparison of lensing and X-ray masses

We can compare the projected mass distribution determined by the lensing analysis to the results from the X-ray data. In Fig. 10 we presented the radial mass distribution in PKS0745 inferred from the X-ray data. Fig. 14 shows this mass profile projected in the plane of the sky, with the lensing masses within a radius of 45.9 kpc overlaid. The results show *good agreement between the lensing and X-ray masses* in the central regions of PKS0745. The agreement between the X-ray mass, $3.2^{+0.8}_{-0.5} \times 10^{13}$ $M_\odot$ within 45.9 kpc (determined under the assumption of spherical symmetry in the ICM) and the lensing mass obtained with the circular model is very good (the lensing mass is $\sim$ 10 per cent lower). For the elliptical lensing models, the lensing mass is $\sim$ 20 per cent lower than the X-ray mass. However, analyses of the X-ray data using ellipsoidal geometries (which unfortunately cannot be firmly constrained by the X-ray image) would lead to mass estimates slightly lower than those from the spherical models, improving the agreement. (The use of a spherically-symmetric geometry in the deprojection analysis of an ellipsoidal cluster leads to an overestimate of the X-ray gas pressure and, therefore, the total gravitating mass as a function of radius.)

Recall (Section 5.2) that the X-ray data only formally constrain the temperature profile (and therefore the mass distribution) of the cluster within $\sim$ 1.5 Mpc. In the 1.5 − 3 Mpc region, the mass distribution has been inferred assuming that the cluster remains isothermal. However, the projected mass within the critical lensing radius is insensitive to uncertainties in the mass distribution at large radii. Truncating the mass distribution at 1.5 Mpc only reduces the projected mass within 50 kpc of the cluster centre by $\sim 5 \times 10^{11}$ $M_\odot$ ($\sim$ 1.4 per cent).

**Table 7.** Results for elliptical mass models

| Model | $r_c$ kpc | Ellipticity 1-b/a | $M_{proj}(r_{arc})$ $10^{13}$ $M_\odot$ | Amplification $\Delta$mag |
|-------|-----------|-------------------|----------------------------------------|---------------------------|
| 1 | 25 | 0.17 | $2.3 - 2.7$ | $2.5 - 4.0$ |
| 2 | 40 | 0.17 | $2.3 - 2.7$ | $2.5 - 4.0$ |
| 3 | 40 | 0.3 | $2.1 - 2.5$ | $2.5 - 4.0$ |

Notes: Mass estimates (within a circular aperture of radius 45.9 kpc) and amplification factors for arc 'A' from the elliptical lensing mass models.

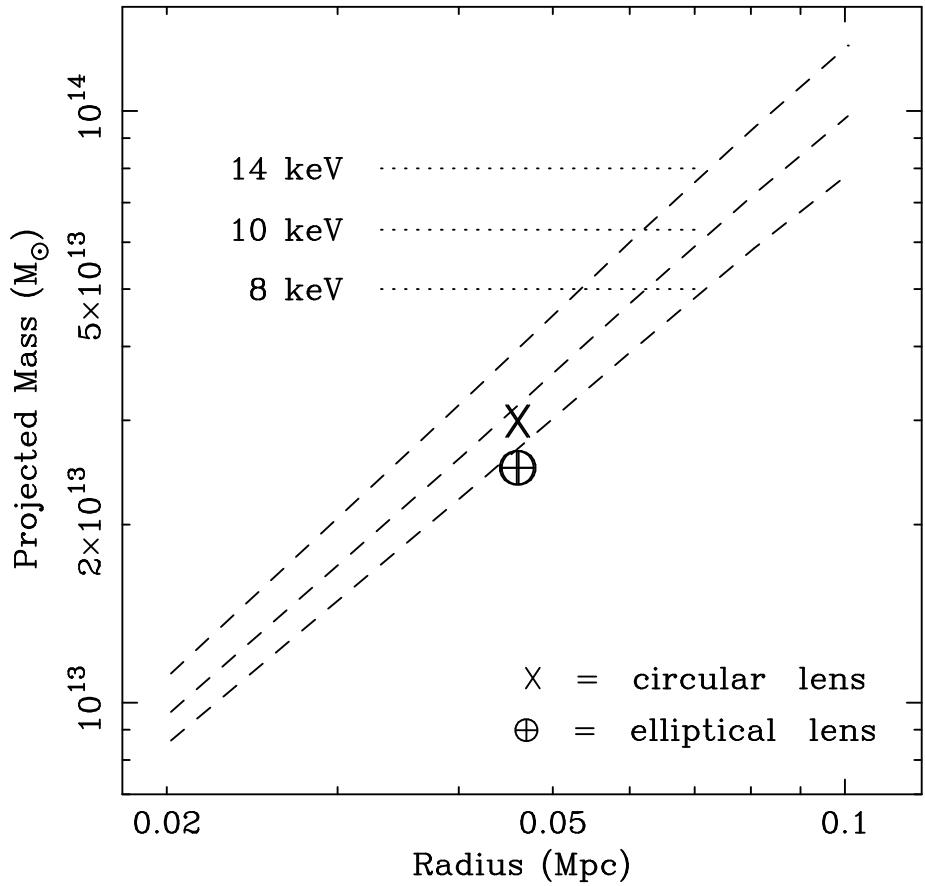

**Figure 14.** A comparison of the X-ray and gravitational lensing mass results. The dashed lines show the projected mass of the cluster determined from the X-ray data (Fig. 10 projected in the plane of the sky). The mass within the critical radius (45.9 kpc) determined from the simple spherical lensing model is marked with a cross. The mass within the same radius, determined from the more sophisticated elliptical lensing analysis, is marked with a circled cross.

## 7   DISCUSSION

PKS0745 is the lowest redshift cluster ($z_{clus} = 0.1028$) in which gravitational lensing effects have been observed to date. Despite its low redshift, however, the discovery of gravitational lensing in PKS0745 is not surprising. PKS0745 is one of the most X-ray luminous (and by implication most-massive) clusters of galaxies known. Its $2 - 10$ keV luminosity of $2.8 \times 10^{45}$ erg s$^{-1}$ (Arnaud *et al.* 1987) is comparable to that of Abell 1689 (Daines *et al.* 1995) and is $\sim 3$ times that of Abell 2218 (McHardy *et al.* 1990; David *et al.* 1993). X-ray observations of PKS0745 then identify it as a strong candidate in which to observe lensing effects. The bright gravitational arc 'A' in PKS0745 is one of the brightest known. If not for its position near the Galactic plane, which hinders optical studies of the cluster, the presence of lensing effects in PKS0745 would probably have been discovered at a much earlier date.

The spectral and imaging data presented here provide a consistent description of the cooling flow in PKS0745. The cooling flow has a mass deposition rate of $\sim 1000$ M$_\odot$ yr$^{-1}$ and the deposition is concentrated within the central $\sim 200$ kpc of the cluster. The SIS data suggest the presence of an intrinsic column density of $\sim 10^{21}$ atom cm$^{-2}$ associated with the flow. Detailed spectroscopy of the bright gravitational arc 'A', at a projected radius of $\sim 50$ kpc in the cluster, might allow absorption features associated with this material to be detected.

The most important result of this paper is the agreement of the mass determinations from the gravitational lensing and X-ray methods (Fig. 14). As discussed in Section 1, studies of the lensing clusters Abell 1689 and Abell 2218 (Daines *et al.* 1995; Kneib *et al.* 1995; Miralda-Escude & Babul 1995) have highlighted discrepancies of $\sim$ a factor two in the X-ray and lensing masses of those systems. These results have led Miralda-Escude & Babul (1995) and Loeb & Mao (1995) to suggest that non-thermal processes, such as magnetic fields and/or large turbulent motions, could contribute significantly to the support of the X-ray gas and invalidate the assumption of hydrostatic equilibrium involved in the X-ray analyses. The agreement of the mass determinations for PKS0745 presented here demonstrate that the X-ray gas in this cluster is in hydrostatic equilibrium. Non-thermal pressure components are not required by the data, and components of the scale required to explain the mass discrepancies in Abell 1689 and Abell 2218 can be firmly ruled out.

The most obvious difference between PKS0745 and Abell 1689 and Abell 2218 is in their morphologies. As discussed in Section 3, the X-ray emission from PKS0745 is smooth, elliptical and sharply peaked onto the CCG. The cluster is regular and relaxed. In contrast, ROSAT HRI observations of Abell 2218 (Kneib *et al.* 1995) show the X-ray emission from the core of the cluster to be highly disturbed, suggesting that the cluster has recently undergone a merger event.

Optical observations of Abell 1689 and Abell 2218 also suggest complex dynamical states in these clusters. Abell 1689 is a Bautz-Morgan Type II-III cluster and Abell 2218 is Bautz-Morgan type II. Both Abell 1689 and Abell 2218 have high galaxy velocity dispersions; $2355^{+238}_{-183}$ km s$^{-1}$ (Teague, Carter & Gray) for Abell 1689 and $1370^{+160}_{-210}$ km s$^{-1}$ (Le Borgne *et al.* 1992) for Abell 2218, indicating substantial substructure (Daines *et al.* 1995). Whereas at optical wavelengths PKS0745 is dominated by a single CCG, Abell 2218 contains two similarly dominant galaxies (Kneib *et al.* 1995) and Abell 1689 has three large galaxies in its core (Daines *et al.* 1995).

The regular, relaxed nature of PKS0745, in contrast to the more dynamically-active states of Abell 1689 and Abell 2218, offers the most likely explanation for the disparity in the X-ray and lensing masses for the latter systems. During the merger of a subcluster with a cluster, the tightly-bound core of the subcluster will rapidly be deposited into the core the larger system (Fabian & Daines 1991). Where the cluster and subcluster mass components overlap along the line of sight, the projected mass density will be enhanced. During the merger the central ICM may *temporarily* leave hydrostatic equilibrium as it is shocked and turbulently mixed. However, when the merger is complete, the cluster gas will relax and return to hydrostatic equilibrium on a timescale of a few sound crossing times.

A merger event, viewed from a particular direction, can temporarily boost the surface mass density of a cluster above the critical density over a wider range of radii than for a relaxed cluster of the same total mass. This enhances the probability of observing gravitational arcs in the cluster and leads to a natural bias for the most-spectacular examples of gravitational lensing to be observed in massive clusters with significant line-of-sight substructure. This is consistent with the observations of Abell 370, Abell 1689, Abell 2218, Cl2244-02 and Cl0024+1654 which are amongst the best examples of lensing clusters and which exhibit complex optical morphologies. Also, as noted by Daines *et al.* (1995), although the dependence of $r_{arc}$ on $z_{clus}$, $z_{arc}$ and $M_{proj}(r_{arc})$ is complex, there is a tendency for arcs in clusters with substructure to be observed at larger radii than arcs in relaxed clusters of similar X-ray luminosity. For example, in the substructured systems Abell 1689, $r_{arc} \sim 200$ kpc (Daines *et al.* 1995), Cl0024+1654, $r_{arc} \sim 200$ kpc (Kassiola, Kovner & Fort 1992) and for Abell 370, $r_{arc} \sim 300$ kpc (Fort *et al.* 1988). In the relaxed systems PKS0745, $r_{arc} \sim 50$ kpc, Abell 963, $r_{arc} \sim 80$ kpc (Lavery & Henry 1988) and for MS2137-23, $r_{arc} \sim 90$ kpc (Mellier *et al.* 1993).

It is relevant to note here that most X-ray luminous clusters of galaxies contain cooling flows (Edge, Stewart & Fabian 1992). Indeed, the natural state of a regular, relaxed cluster appears to be with a cooling flow in its core. Simulations of cluster/subcluster mergers (McGlynn & Fabian 1984) show that once a cooling flow is established, only the merger of the host cluster with a similarly-sized subcluster is likely to re-heat the central ICM to the extent that the cooling flow is 'turned off'. The highly disturbed X-ray morphology and absence of a cooling flow in Abell 2218 (established from a deprojection analysis of the ROSAT HRI data) is then a good indicator of merger activity in this cluster. This argument also suggests that other lensing clusters without cooling flows are likely to contain substantial substructure in their mass distributions and exhibit a discrepancy between their X-ray and lensing masses.

Finally, it is important to recognise the need for using physically-appropriate models when determining the mass of clusters from X-ray data. For clusters with large cooling flows, in which the central ICM is substantially multiphase, the use of single-phase models can lead to significant underestimates of the central mass (or overestimates of the gravitational core radius). For PKS0745, a spectral analysis of the central region of the cluster with a single-phase model implies a gas temperature of $\sim 6$ keV and a projected mass within the lensing radius of $\sim 10^{13}$ M$_\odot$. In contrast, the use of a multiphase cooling-flow model (which is statistically required by the data) implies a central temperature for the ambient cluster gas of $\sim 10$ keV and a mass within the lensing radius of $\sim 3 \times 10^{13}$ M$_\odot$.

## 8  CONCLUSIONS

This paper has presented a consistent determination of the distribution of mass in PKS0745-191 using both X-ray and gravitational lensing methods. We have highlighted the importance of adopting a physically-appropriate multiphase model in the analysis of the X-ray data. The agreement of the X-ray and lensing mass results for PKS0745 confirms that the ICM is in hydrostatic equilibrium with the gravitational potential of the cluster. The agreement of the X-ray and lensing masses for PKS0745 contrasts with the results for Abell 1689 and Abell 2218, where a discrepancy of $\sim$ a factor 2 is observed. We have suggested that these discrepancies are due to line-of-sight enhancements in the surface mass densities, due to ongoing merger events. This conclusion is supported by the substructure observed in optical and X-ray data for the clusters.

PKS0745 is the lowest-redshift cluster in which gravitational lensing effects have so far been observed. We have shown that the properties of the brightest lensed source in the cluster are consistent with those of an early-type spiral galaxy, at a redshift of 0.433, with ongoing star formation. The X-ray spectra and images of PKS0745 confirm that the cluster contains one of the largest known cooling flows, with an integrated mass deposition rate of $\sim 1000$ $M_\odot$ yr$^{-1}$.

### ACKNOWLEDGMENTS

We are grateful to H. Richer, G. Fahlman and M. Bremer for their help with the optical observations. We thank A. Edge and particularly R. Johnstone for discussions, and D. White for coding used in the analysis of the X-ray data. SWA acknowledges receipt of a PPARC postdoctoral fellowship. ACF thanks the Royal Society for support.